\documentclass[usenatbib]{mnras}

\usepackage[T1]{fontenc}
\usepackage{ae,aecompl}

\usepackage{graphicx}	
\usepackage{multirow}
\usepackage{pdflscape}
\usepackage{amssymb}
\usepackage{amsmath}	
\usepackage{bm}
\usepackage{morefloats}
\usepackage{txfonts}
\usepackage{hyperref}
\usepackage{natbib}
\usepackage{threeparttable}
\usepackage{qtree}
\usepackage{dirtree}

\newcommand{\hvc}{HVC}
\newcommand{\hvcs}{HVCs}

\newcommand{\ism}{ISM}
\newcommand{\him}{HIM}
\newcommand{\cgm}{CGM}

\newcommand{\rcloud}{R_{\rm cl}}
\newcommand{\mcloud}{M_{\rm cl}}
\newcommand{\tcloud}{T_{\rm cl}}

\newcommand{\mucloud}{\mu_{\rm cl}}

\newcommand{\tamb}{T_{\rm amb}}
\newcommand{\namb}{n_{\rm amb}}

\newcommand{\mmax}{M_{\rm BE}}

\newcommand{\cs}{c_{\rm s}}
\newcommand{\csbar}{\bar{c}_{\rm s}}
\newcommand{\lf}{\lambda_{\rm F}}
\newcommand{\lel}{\lambda_{\rm e}}
\newcommand{\LT}{L_{\rm T}}
\newcommand{\kms}{\,\textrm{km}\,\textrm{s}^{-1}}
\newcommand{\zsolar}{\,\textrm{Z}_\odot}
\newcommand{\msolar}{\,\textrm{M}_\odot}

\newcommand{\ccm}{\,\textrm{cm}^{-3}}
\newcommand{\snapone}{$25~$Myr}
\newcommand{\snaptwo}{$50~$Myr}
\newcommand{\snapthree}{$75~$Myr}
\newcommand{\snapfour}{$100~$Myr}
\newcommand{\tsc}{\tau_{\rm sc}}

\title[Thermal conduction in resting multiphase clouds]{The effect of saturated thermal conduction on clouds in a hot plasma}

\author[B. Sander \& G. Hensler]{
Bastian Sander,$^{1,2}$\thanks{E-mail: bastian.sander@hs-anhalt.de, gerhard.hensler@univie.ac.at}
Gerhard Hensler,$^{1}$\footnotemark[1]
\\
$^{1}$Department of Astrophysics, University of Vienna, T\"urkenschanzstra\ss e 17, A-1180 Vienna, Austria\\
$^{2}$Anhalt University of Applied Sciences, Strenzfelder Allee 28, 06406 Bernburg, Germany
}

\date{Accepted XXX. Received YYY; in original form ZZZ}

\pubyear{2021}

\begin{document}
\label{firstpage}
\pagerange{\pageref{firstpage}--\pageref{lastpage}}
\maketitle

\begin{abstract}
We numerically investigate the internal evolution of multiphase clouds, which are at rest with respect to an ambient, highly ionized medium (\him) representing the hot component of the circumgalactic medium (\cgm).
%
Time-dependent saturated thermal conduction and its implications like condensation rates and mixing efficiency are assessed in multiphase clouds.
%
Our simulations are carried out by using the adaptive mesh refinement code {\sc Flash}. We perform a grid of models of which we present here those characteristic for the presented study. The model clouds are initially in both hydrostatic and thermal equilibrium and are in pressure balance with the \him. Thus, they have steep gradients in both temperature and density at the interface to \him{} leading to non-negligible thermal conduction. Several physical processes are considered numerically or semi-analytically: thermal conduction, radiative cooling and external heating of gas, self-gravity, mass diffusion, and dissociation of molecules and ionization of atoms.
%
It turns out that saturated thermal conduction triggers a continuous condensation irrespective of cloud mass. Dynamical interactions with ambient \him{} all relate to the radial density gradient in the clouds: (1) mass flux due to condensation is the higher the more homogeneous the clouds are; (2) mixing of condensed gas with cloud gas is easier in low-mass clouds, because of their shallower radial density gradient; thus (3) accreted gas is distributed more efficiently. A distinct and sub-structured transition zone forms at the interface between cloud and \him, which starts at smaller radii and is much narrower as deduced from analytical theory.
\end{abstract}




\section{Introduction}\label{sec:intro}

Astrophysical environments on and below galactic scales are inhomogeneous and not isothermal, which is observed in the interstellar medium \citep[\ism,][]{05lequeux,09shelton,21carraro} and circumgalactic medium \citep[\cgm,][]{12putmanpeekjoung,17tumlinsonpeepleswerk}. Sample mechanisms that drive the Galactic matter cycle by maintaining these different physical states (so-called \emph{phases}) on diverse spatial scales are galactic outflows \citep[][]{89normanikeuchi,03mccluregriffithsetal,11romeeloppenheimerfinlator} and accretion of gas \citep[][]{14fraternali,16richter,17foxdave}, supernovae \citep[][]{11hensler}, and stellar winds \citep[][]{18zhang}. The coexistence of these phases in the \ism{} and \cgm{} implies non-equilibria situations being continuously triggered. These non-equilibria situations lead to various dynamic processes like hot accretion flows \citep[][]{12faghei}, plasma cooling \citep[][]{06lehnersavagewakker}, or thermal conduction \citep[][]{04nipotibinney,04pittardetal,16brueggenscannapieco}, such that the phases in the \ism{} and \cgm{} do not only contact and envelope each other, but also penetrate into each other in order to equilibrate by exchange of energy and mass. A static situation of the \ism{} is represented by the theory of a two-phase medium \citep[][]{69fieldgoldsmithhabing}, which is later extended by a third hot phase \citep[][]{77mckeeostriker}. Recent studies and observations reveal a more complex dynamic picture of the hot haloes around galaxies \citep[][]{12breitschwerdtetal,14falcetagoncalvesetal,19leelee,22damleetal}. High pressure and gravity in the cold neutral matter (CNM) balance with low pressure in the highly ionized medium (\him) at their common interface, such that steep gradients in both density and temperatures are present. They in turn lead to a flow of heat and matter towards the CNM. Analytical approaches propose the evaporation of clouds in a few dynamical times \citep[][]{82balbusmckee,84giuliani,07nagashimainutsukakoyama}, which contradicts observations of cool clouds embedded in diverse hot environments. Thus, the clouds are able to resist evaporation by combined physical mechanisms. 

Previous numerical studies of multiphase clouds considering plasma heating and cooling, self-gravity, mass diffusion, and thermal conduction reveal that thermal conduction is saturated and leads to condensation thus providing a mechanism to mix ambient gas into clouds and, by this, enhancing their metallicity \citep[two-dimensional simulations by][hereafter VH07b]{07vieserhensler2}. In a follow-up study \citet[][]{07vieserhensler1} show that the lifetime of moving clouds is distinctly extended by thermal conduction, which efficiently suppresses Kelvin-Helmholtz instabilities. \citet[][]{19sanderhensler} work out that (1) self-gravity is important for even sub-Jeans clouds, and (2) thermal conduction may not be neglected for gaseous spheres even in presence of strong magnetic fields. Very recently, \citet[][hereafter SH21]{21sanderhensler} analyze the evolution of high-velocity clouds (\hvcs) in a hot \cgm{} and figure out that thermal conduction together with deceleration by drag naturally leads to the suppression of star formation in compact \hvcs, which conforms with observations \citep[e.g.,][]{02daviesetal,07hoppschulteladbeckkerp}. In the present work we extend these investigations by analyzing saturated thermal conduction in an isolated manner whithout being superimposed by hydrodynamic interactions.

This paper is structured as follows: in Sections~\ref{sec:heatcond} and \ref{sec:physprocs} we describe the physical processes being considered in our simulations, Section~\ref{sec:models} contains the applied numerical code and the model setup. The model clouds and the results of the simulations are discussed in Section~\ref{sec:discussion}. The paper is summarized in Section~\ref{sec:sumcon}, where also the major conclusions are drawn.

\section{Thermal conduction}\label{sec:heatcond}
In the presence of large temperature gradients the classical description of the heat flux given by \citet{62spitzer},
\begin{equation}
\bm{q}_{\rm class}=-\kappa_{\rm class}\nabla T,\label{equ:satcond-1}
\end{equation}
with the coefficient for thermal conduction $\kappa_{\rm class}=1.84\times 10^{-5}T^{5/2}/\ln(\Omega)$, and the Coulomb logarithm $\ln(\Omega)=29.7+\ln[T/(10^6\sqrt{n_{\rm e}})]$, yields a too high amount of energy being conducted by electrons \citep[e.g.,][]{84campbell}. Instead, a saturated heat flux $\bm{q}_{\rm sat}$ \citep[see][]{77cowiemckee,77mckeecowie} accounts for a finite reservoir of electrons and thus converges to a maximum value if all electrons are conducting heat irrespective of the steepness of the temperature gradient. It has been experimentally verified by \citet{80graykilkenny} that
\begin{equation}
\bm{q}=\left\{
\begin{array}{ll}
\bm{q}_{\rm class} & ,\;\lel/\LT\lesssim 2\times 10^{-3} \\
\bm{q}_{\rm sat} & ,\;{\rm else}
\end{array} .
\right.\label{equ:satcond-2}
\end{equation}
The first essential spatial scale is the mean free path of electrons in the hot, ambient medium,
\begin{equation}
\lel=t_{\rm e}\sqrt{\frac{3k\tamb}{m_{\rm e}}},\label{equ:parameter-2}
\end{equation}
with temperature $\tamb$ and electron-electron equipartition time \citep{62spitzer}
\begin{equation}
t_{\rm e}=\frac{3\sqrt{m_{\rm e}}(k\tamb)^{3/2}}{4\sqrt{\pi}n_{\rm e}e^4\ln(\Omega)}.\label{equ:parameter-3}
\end{equation}
The second spatial scale is the local temperature scale-height, $\LT=T/|\nabla T|$, which provides the typical spatial scale for temperature variation. If $\lel$ is significantly less than $\LT$ (diffusion limit), temperature gradients are levelled off and the classical approach for thermal conduction is valid. In the opposite case, even strong temperature variations are preserved (free-streaming limit) and thermal conduction becomes saturated. \citet[][hereafter CM77]{77cowiemckee} suggest the expression 
\begin{equation}
\bm{q}_{\rm sat}=5\Phi\varrho \cs^3\label{equ:satcond-9}
\end{equation}
for the saturated heat flux, which depends on the speed of sound $\cs$, local density $\varrho$, and a factor $\Phi$ accounting for the uncertainties in the definition of (\ref{equ:satcond-9}) regarding contributions in a plasma by magnetic fields, currents, instabilities etc. It holds $\Phi=1.1$ if both ion and electron temperatures are equal. We use $\Phi=1$ throughout all of our simulations analyzed in this work.

To ensure continuity at the transition between $\bm{q}_{\rm class}$ and $\bm{q}_{\rm sat}$ a proper representation for the heat flux must be chosen. Within this work the expression of \citet{92slavincox}
\begin{equation}
\bm{q}_{\rm eff}=\bm{q}_{\rm sat}\left(1-\exp\left\{-\frac{|\bm{q}_{\rm class}|}{|\bm{q}_{\rm sat}|}\right\}\right)\label{equ:satcond-3}
\end{equation}
is used. It correctly converges to either of the heat fluxes in equation~(\ref{equ:satcond-2}) for very small or very large temperature gradients, respectively.

The conductivity according to the effective heat flux (\ref{equ:satcond-3}) reads
\begin{equation}
\kappa_{\rm eff}=\kappa_{\rm sat}\left(1-\exp\left\{-\frac{\kappa_{\rm class}}{\kappa_{\rm sat}}\right\}\right),\label{equ:satcond-5}
\end{equation}
with $\kappa_{\rm sat}=5\varrho \cs^3/|\nabla T|$. We follow the procedure described in SH21 for calculating the temperature update according to saturated thermal conduction. Namely, we use the semi-implicit Crank-Nicolson method \citep[][]{47cranknicolsonhartree}, which is applied in a directionally split manner, i.e. the temperature is updated along one spatial dimension while being kept constant along the other two dimensions. This procedure is repeated for the other dimensions, too. The order of direction sweeps is permuted analogously to the hydrodynamics (cf. SH21). From the new temperature distribution an adjusted heat flux 
\begin{equation}
\bm{q}_{\rm eff}=-\kappa_{\rm eff}\nabla T\label{equ:satcond-7}
\end{equation}
finally accounts for the change in energy by thermal conduction. It has to be noted that the temperature gradient can be very steep, especially across the cloud surface. The heat flux (\ref{equ:satcond-7}) can thus be very high in the vicinity of regions of nearly zero heat flux. We account for this special situation by a slope limiter (cf. SH21).

\citet[][]{90begelmanmckee} introduce the Field length
\begin{equation}
\lf=\sqrt{\frac{\kappa_{\rm amb}T_{\rm amb}}{n_{\rm amb}^2\Lambda_{\rm max}}},\label{equ:resolution-1}
\end{equation}
as a characteristic spatial scale to distinguish between growth and suppression of thermal instabilities. The Field length relates the thermal conductivity $\kappa_{\rm amb}$ with cooling efficiency $\Lambda_{\rm max}=\max\{\Lambda_0,\Gamma_0/\namb\}$ in the surrounding, hot phase that directly contacts the cloud. The rate coefficients for radiative cooling, $\Lambda_0$ [erg cm$^{-3}$ s$^{-1}$], and heating, $\Gamma_0$ [erg s$^{-1}$], denote the respective density-independent powers being radiated \citep[][]{76raymondcoxsmith,89boehringerhensler}. Thus, by equation~(\ref{equ:resolution-1}) the Field length of ambient gas is calculated. If gas phases change, then $\lf$ changes accordingly. So, $\lf$ is a function of radius and time in cool clouds embedded in a hot plasma with time-dependent thermal conduction. If the cloud is interpreted as a cool density inhomogeneity embedded in the hot phase, its radius can be compared to $\lf$ and a direct consequence for the thermal structure of the cloud can be derived: if $\rcloud\gg\lf$, the cloud is dominated by external heating and radiative cooling while for $\rcloud\ll\lf$ classical thermal conduction governs. If, however, $\rcloud\sim\lf$, saturated thermal conduction is the dominating effect. The Field length must be resolved by at least three grid cells in numerical simulations \citep[][]{04koyamainutsuka,09gressel} to prevent results from being dependent on spatial resolution. We account for this in our simulations by resolving $\lf$ in equation~(\ref{equ:resolution-1}) by $206$ cells in the massive model clouds and $275$ cells in the low-mass clouds.

\begin{figure}
\centering
\includegraphics[width=.5\textwidth]{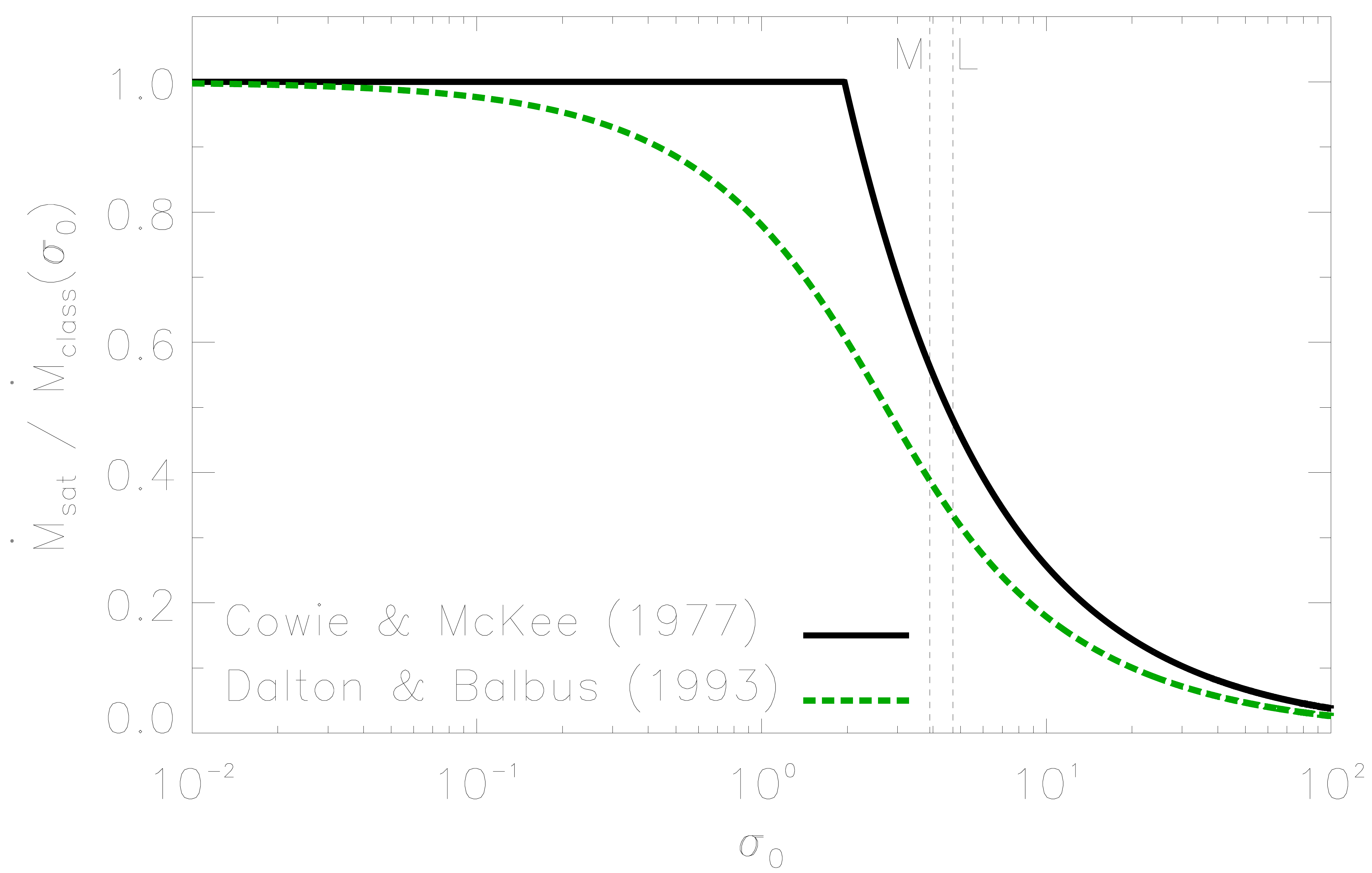}
\caption{Ratios of saturated mass-loss rate to classical mass-loss rate analytically calculated by CM77 (\emph{black solid line}) and \citet[][]{93daltonbalbus} (\emph{green dashed line}). The \emph{vertical thin dashed lines} mark the values for $\sigma_0$ used for the simulated cloud models M and L (cf. Table~\ref{table:models-3}).}
\label{fig:parameter-1}
\end{figure}

CM77 introduce a global saturation parameter
\begin{equation}
\sigma_0=\frac{\left(\tamb/1.54\times 10^7{\rm [K]}\right)^2}{\namb\Phi \rcloud{\rm [pc]}},\label{equ:parameter-1}
\end{equation}
which can be used to discriminate between classical ($\sigma_0<1$) and saturated ($\sigma_0>1)$ thermal conduction. Here, $\rcloud$ is the cloud radius in terms of parsec, and $\Phi$ is defined in equation~(\ref{equ:satcond-9}). The mass loss is considerably reduced with increasing $\sigma_0$ (see solid line in Fig.~\ref{fig:parameter-1}). For $\sigma_0\gg 1$ also viscous heating becomes relevant and thermal conduction is best treated by a two-fluid approach \citep[][]{82balbusmckee}. Based on (\ref{equ:parameter-1}) CM77 determine the rate of evaporation due to saturated thermal conduction by
\begin{equation}
\dot{M}_{\rm CM}=3.25\times 10^{18}\namb\sqrt{\tamb}\rcloud^2{\rm [pc]}\Phi F(\sigma_0)\;{\rm g\;s^{-1}}\label{equ:satcond-10}
\end{equation}
for clouds without self-gravity, where
\begin{equation}
F(\sigma_0)=2\left[(\sigma_0H)^{1+Ma^2}\exp\left\{-2.5Ma^2\right\}\right]^{1/(6+Ma^2)},\label{equ:satcond-11}
\end{equation}
and 
\begin{equation}
H=\left\{
\begin{array}{ll}
h^h/(h-1)^{h-1} &,Ma\leq 1 \\
11.5 &,Ma> 1
\end{array}
\right.\quad,\quad h=\frac{11+Ma^2}{1+Ma^2},\label{equ:satcond-12}
\end{equation}
and $Ma$ is the Mach number of the flow around the cloud or, more specific, within the zone of saturated heat flux. \citet[][]{93daltonbalbus} extend these investigations by introducing a continuous transition between classical and saturated heat flux. By simultaneously solving their equations (25) and (30) they obtain the ratio $\omega(\sigma_0)$ (cf. dashed line in Fig.~\ref{fig:parameter-1}) between saturated and classical mass-loss rate. Hence,
\begin{equation}
\dot{M}_{\rm DB}=\omega(\sigma_0)\dot{M}_{\rm class}=\omega(\sigma_0)\frac{16\pi\mu_{\rm\him}\kappa_{\rm\him}\rcloud}{25k_{\rm B}}.
\label{equ:satcond-13}
\end{equation}
Theoretically, gas should evaporate from clouds for $\lel\lesssim \rcloud\ll\lf$ and condense onto them if $\rcloud\gg\lel$ and $\rcloud\gtrsim\lf$. \citet{90mckeebegelman} show that hot material condenses on isolated clouds if $\rcloud\sim 0.24\ldots 0.36\lf$ (see also VH07b).

\section{Physical processes}\label{sec:physprocs}

Beside thermal conduction we cover in our simulations different physical processes that most likely take place in interstellar clouds. They are explained in detail in SH21 thus we summarize only the most important facts here.

\emph{Self-gravity:} The Poisson's Equation is solved in the entire computational domain. It is shown in \citet[][]{19sanderhensler} that self-gravity may not be neglected even in clouds with masses below their Bonnor-Ebert mass. If self-gravity is stronger than acceleration of gas due to evaporation, the net acceleration at cloud surface points in the same direction like the density gradient. Hence, the condition for Rayleigh-Taylor (RT) instability is not given. So, it is expected that self-gravity significantly suppresses RT instabilities at the cloud surface \citep[][]{93murrayetal}.

\emph{Dissociation and ionization:} The simulations contain cool clouds and hot plasma, thus temperatures range from a few hundreds up to several $10^6~$K. While at lowest possible temperatures molecules can exist, they successively dissociate and finally become totally ionized with rising temperature. In our simulations we semi-analytically calculate both dissociation and ionization for hydrogen only. The fraction of \ion{H}{ii} is then related to He and metals. So, we have multiple phases (molecular, atomic, ionized) simultaneously present in our clouds. We note that we do not trace single species but rather have one fluid for which the composition is calculated based upon a given metallicity. The fractions of dissociation and ionization affect the magnitudes of cooling rates.

\emph{Plasma heating and cooling:} The rates of heating, $\Gamma$, and cooling, $\Lambda$, of the plasma are treated semi-analytically. Depending on temperature we account for molecular line cooling \citep{85falgaronepuget}, atomic line cooling \citep{72dalgarnomccray}, and bremsstrahlung \citep{89boehringerhensler}. The plasma is heated owing to the photoelectric effect on dust particles \citep{06weingartnerdrainebarr}, ionization by UV radiation, by X-rays, and by cosmic rays \citep[][]{03wolfireetal}, thermalization of turbulent motions, and condensation of molecular hydrogen on dust particles \citep[both in][]{10tielens}.

Both $\Gamma$ and $\Lambda$ depend on local values of metallicity, temperature, density, and the fractions of dissociation and ionization. If metallicity is high, cooling rates are stronger and hence the temperature is lower in thermal equilibrium, which implies a lower pressure. Consequently, especially the mixing regions in the outskirts of clouds are expected to have a decreased pressure, which results in a flow of matter towards the outskirts.

\emph{Mass diffusion:} We consider the transport of matter by diffusion using the formulation of the Chapman-Enskog expansion, which is valid at low gas densities \citep{60birdstewartlightfoot}.

\emph{Simplifications:} Radiative transport is neglected, because we assume the model clouds to be optically thin. Magnetic fields are neglected as even strong dipole fields do not remarkably suppress the heat flux integrated over the entire cloud surface \citep[][]{19sanderhensler}. It has been shown by \citet[][cf. equation~11 there]{89tribble} that tangled magnetic field lines suppress the effective thermal conductivity between adjacent gas phases at different temperatures only by some small amount. A similar result is reported by \citet[][]{01malyshkin} for a negligible homogeneous component of the magnetic field \citep[see also][]{95tao}.

We further consider chemodynamics on a simplified level, i.e. different species are not explicitly traced by their respective conservation laws. We rather calculate the fractions of H$_2$, \ion{H}{i}, \ion{H}{ii}, He, electrons, and metals in a semi-analytical fashion based on the local fractions of dissociation and ionization, respectively, for a single fluid.

\emph{Hydrodynamics:} We solve the Euler Equations for clouds resting in the \him{} and close them by an equation of state for the ideal gas. According to the source terms of self-gravity, heating and cooling, and thermal conduction, the hydrodynamics conservation law for total energy, $e_{\rm tot}$, we solve numerically is given by
\begin{equation}
\frac{\partial\varrho e_{\rm tot}}{\partial t}+\nabla\left[(\varrho e_{\rm tot}+P)\bm{v}\right]=\varrho\bm{v}\bm{g}+\Gamma-\Lambda-\nabla\bm{q}_{\rm eff}.\label{equ:satcond-8}
\end{equation}

\section{The simulations}\label{sec:models}

We conduct full 3D hydrodynamics simulations of cool, stratified, dense, multiphase clouds that are at rest with respect to an ambient, hot, rarefied medium resembling the \him{} by means of the publicly available {\sc Flash} code\footnote{see \href{http://flash.uchicago.edu/site/flashcode/}{http://flash.uchicago.edu/site/flashcode/}} \citep[][]{00fryxelletal,09dubeyetal}. The (inviscid) Euler Equations are solved on a Cartesian, adaptively refined grid by applying the piecewise parabolic method \citep[PPM,][]{84collelawoodward,84woodwardcolella}. The Poisson's Equation is solved by a Multigrid solver, which is able to deal with arbitrary mass distributions. See SH21 for further details on the {\sc Flash} code and how we handle steep temperature gradients.

The present work investigates in more detail the effect of saturated thermal conduction on multiphase clouds and compares to previous numerical (VH07b) and analytical \citep[CM77,][]{77mckeecowie,82balbusmckee,84giuliani,90begelmanmckee,90mckeebegelman,07nagashimainutsukakoyama} works. For this purpose cool, molecular to low-ionized (i.e. multiphase), dense, stratified clouds are set up at rest in a hot, tenuous, fully ionized plasma resembling the highly ionized medium (\him, $T\gtrsim 10^{5.5}$K) in the \cgm. Within this paper we focus on the meaning of thermal conduction for the evolution of interstellar clouds that are investigated in SH21. The present study serves for investigating the non-dynamical processes that provide the inital conditions for \hvcs. In particular, we are interested in the stabilizing effect thermal conduction has on \hvcs{} by (i) formation of a transition zone between cloud and ambient gas, and (ii) mass gain due to condensation of hot gas onto the cloud. For that purpose, our model clouds are setup at rest with respect to the ambient hot gas, which means they are neither in translational nor rotational motion. Dynamical effects like ram-pressure stripping or the Bernoulli effect hence do not superpose diffusion and we are able to study thermal conduction in the clouds in a fundamental manner.

The clouds divide into two mass categories: massive clouds (category \emph{M}) and low-mass clouds (category \emph{L}). Each of the model clouds considers all physical processes described in Section~\ref{sec:physprocs}. The reference clouds (M, L) additionally account for thermal conduction while the model variations M\_nc and L\_nc do not, i.e. there is no conductive input of thermal energy from the ambient medium.

Initially, all four model clouds are in both hydrostatic and thermal equilibrium. By that, the radial temperature profile of the cloud is governed by 
\begin{equation}
\frac{\partial T}{\partial r}=-\varrho(T)\frac{GM(r<\rcloud)}{r^2}\left\{\frac{\partial}{\partial T}\left[\frac{\varrho(T)kT}{\mu(\varrho,T)}\right]\right\}^{-1},\label{equ:initial-2}
\end{equation}
where $\mu(\varrho,T)$ is the mean molecular weight. From equation~(\ref{equ:initial-2}) the profiles of density and pressure are derived (see fig.~2 in SH21) and the radius $\rcloud$ is obtained at a cloud-centric distance, where $P_{\rm cl}(r=\rcloud)$ equals the pressure of the ambient medium, $P_{\rm amb}$. As being discussed in the present paper, model clouds of equal mass differ only in an additional heat input by thermal conduction, which destroys thermal balance over time. The Poisson's Equation is solved in the entire computational domain, hence \him{} is accreted by the clouds due to gravity. The non-conductive model clouds thus serve as a testbed of accretion by gravity compared to condensation by thermal conduction. Moreover, initial hydrostatic and thermal equilibrium must be maintained in clouds without thermal conduction, so they prove the implemented numerics to work correctly.

According to the \hvc{} models in SH21 we still distinguish the metal content in the gas phases: All model clouds have an initial metallicity of $0.1\zsolar$ and the \him{} has solar metallicity. By using different metallicities we are able to trace the enrichment of the cloud by heavy elements from \him{}, which is triggered by gas mixing due to thermal conduction. As already mentioned in Section~\ref{sec:physprocs}, our treatment of species implies that there is no effect of metallicity on gas dynamics. Temperature and density of the \him{} in our simulations resemble those for the circumgalactic medium \citep[][]{17tumlinsonpeepleswerk}.
\begin{table*}
\caption{Simulated model clouds with (models M, L) and without (models M\_nc, L\_nc) thermal conduction (also indicated in \emph{column 3} by $+$ or $-$). \emph{Columns 4} to \emph{9} consecutively list initial values of radius, mass, saturation parameter (cf. equation~(\ref{equ:parameter-1})), ratio of mean free path of electrons to scale height of temperature (cf. equation~(\ref{equ:satcond-2})), ratio of cloud radius to Field length, and the cloud mass in terms of Bonnor-Ebert mass. \emph{Columns 10} and \emph{11} tabulate the side length of the computational cube and the finest numerical resolution, respectively, and \emph{column 12} shows the resolution of the Field length. The initial central values for temperature and gas density are given in \emph{columns 13} and \emph{14}. The metallicity in all model clouds is initially $0.1\zsolar$.}
\label{table:models-3}
\centering
\begin{threeparttable}
\begin{tabular}{c l c c c c c c c c c c c c}
\hline
Model & Model & thermal & $\rcloud$ & $\mcloud$ & \multirow{2}{*}{$\sigma_0$\tnote{(*)}} & \multirow{2}{*}{$\lel/\LT$\tnote{(*)}} & \multirow{2}{*}{$\rcloud/\lf$\tnote{(*)}} & \multirow{2}{*}{$\mcloud/\mmax$} & $L$ & $\Delta x$ & \multirow{2}{*}{$\lf/\Delta x$\tnote{(*)}} & $T_{\rm centre}$ & $\varrho_{\rm centre}$ \\ 
group & name & cond. & [pc] & [$10^4\msolar$] & & & & & [pc] & [pc] & & [K] & $\left[10^{-21}\right.$g cm$\left.^{-3}\right]$ \\ 
(1) & (2) & (3) & (4) & (5) & (6) & (7) & (8) & (9) & (10) & (11) & (12) & (13) & (14) \\
\hline
massive & M & $+$ & \multirow{2}{*}{$49$} & \multirow{2}{*}{$9.0$} & $3.8$ & $1.22$ & $0.23$ & \multirow{2}{*}{$0.31$} & \multirow{2}{*}{$260$} & \multirow{2}{*}{$1.02$} & $206$ & \multirow{2}{*}{$265$} & \multirow{2}{*}{$0.8$} \\
clouds & M\_nc & $-$ & & & $-$ & $-$ & $-$ & & & & $-$ & & \\
\hline
low-mass & L & $+$ & \multirow{2}{*}{$28$} & \multirow{2}{*}{$2.2$} & $4.6$ & $0.69$ & $0.16$ & \multirow{2}{*}{$0.09$} & \multirow{2}{*}{$160$} & \multirow{2}{*}{$0.63$} & $275$ & \multirow{2}{*}{$1,160$} & \multirow{2}{*}{$0.02$} \\
clouds & L\_nc & $-$ & & & $-$ & $-$ & $-$ & & & & $-$ & & \\
\hline
\end{tabular}
\begin{tablenotes}
\footnotesize
\item[(*)] This parameter relates to thermal conduction.
\end{tablenotes}
\end{threeparttable}
\end{table*}
By subsuming the parameter ranges for $\sigma_0$, $\lf$, $\lel$, $\LT$ (cf. Section~\ref{sec:physprocs}), and the cloud mass, $\mcloud$, the clouds we are interested in are located in the following subset of the relevant parameter space
\begin{equation}
\sigma_0 > 1\;,\;\rcloud\lesssim\lf\;,\;\mcloud<\mmax,\label{equ:parameter-6}
\end{equation}
whereas the \him{} must satisfy (cf. equation~(\ref{equ:satcond-2}))
\begin{equation}
\lel\gtrsim \LT.\label{equ:parameter-7}
\end{equation}
Table~\ref{table:models-3} contains the realisations of our setups in the parameter spaces (\ref{equ:parameter-6}) and (\ref{equ:parameter-7}). 

The size of the computational volume is $(260~$pc$)^3$ for massive and $(160~$pc$)^3$ for low-mass clouds. Thus, the numerical domain is three times larger in diameter than the cloud, and boundary effects are not observed to have a feedback on the dynamics around the cloud. The chosen size of the integration domain implies a grid spacing of $\Delta x=1.02~$pc ($\Delta x=0.63~$pc) on six levels of refinement for the simulations of massive (low-mass) clouds. In order to prove that obtained results are independent on resolution, we perform additional fine-grained simulations on seven levels of refinement yielding $\Delta x=0.51~$pc ($\Delta x=0.31~$pc) for massive (low-mass) clouds (see appendix \ref{app:resolution}).

\subsection{Massive model clouds}\label{subsec:massiveclouds}
Massive clouds have masses of $9.0\times 10^4\msolar$ that correspond to $31~$per~cent of their Bonnor-Ebert mass \citep{55ebert,56bonnor}
\begin{equation}
\mmax=1.18\left(\frac{k\bar{T}_{\rm cl}}{\bar{\mu}_{\rm cl}}\right)^2G^{-3/2}P_{\rm amb}^{-1/2},\label{equ:initial-3}
\end{equation}
which provides an upper bound for the mass of self-gravitating clouds embedded in a medium with pressure $P_{\rm amb}$ in order to be stable against gravitational instability. The identitiy (\ref{equ:initial-3}) strictly holds for isothermal clouds only. Hence, we use the density-weighted average values of cloud temperature, $\tcloud$, and mean molecular weight, $\mucloud$. The massive models show a distinct core-halo structure in density, temperature and, consequently, in pressure. The central region is dominated by the molecular phase which turns into an extended atomic phase at larger radii. Only a small rim at the boundary consists of warm, slightly ionized gas. Massive clouds are located in a hot plasma with temperature $T_{\rm\him}=5.6\times 10^6~$K and particle density $n_{\rm\him}=0.7\times 10^{-3}\ccm$. They exhibit a metallicity of $0.1$ solar, which leads to a radius of $49~$pc in consideration of equilibrium between cooling and heating in the cloud, and pressure balance between cloud and \him. The central temperature is $265~$K, which quickly raises to $\sim 970~$K at $10~$pc and moderately increases to $\sim 1,650~$K at the boundary (Fig.~\ref{fig:both-temp_diss}, upper plot). That is, the molecular phase is confined to the inner $10~$pc, beyond which dissociation is complete and only the atomic phase is present. Because temperature and hence conductivity are initially moderate inside the cloud, classical thermal conduction takes place. Since $\kappa_{\rm class}\propto T^{5/2}$ a substantial heat transport towards the central cloud region is expected especially around $10~$pc by pure thermodynamics. However, a present radial density gradient with the steepest slope inside of $15~$pc (Fig.~\ref{fig:both-temp_diss}, lower plot) may hinder a mass flux into the centre.

In model M\_nc thermal conduction is neglected but all other physical processes are identical to model M. Hence, the initial model is identical to the reference cloud in the distribution of each variable (like cloud mass, temperature, density, pressure).

\subsection{Low-mass model clouds}\label{subsec:lowmassclouds}
\begin{figure}
\centering
\includegraphics[width=\linewidth]{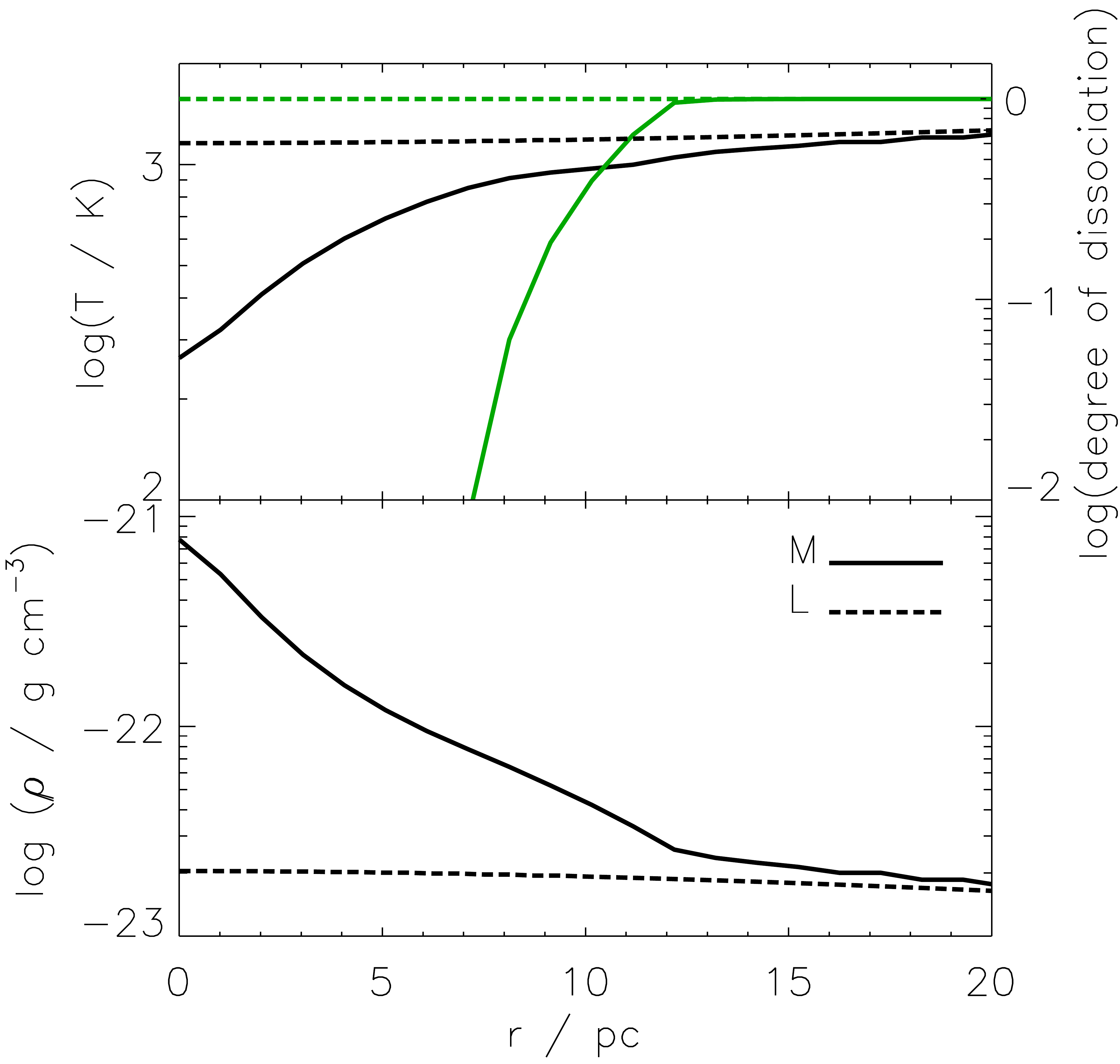}
\caption{\emph{Upper plot:} Initial temperature profile in the central cloud region for models M (\emph{black solid line}) and L (\emph{black dashed line}). The degrees of dissociation (\emph{green lines}) are shown for each model cloud. \emph{Lower plot:} Initial radial profiles of gas density for both models.}
\label{fig:both-temp_diss}
\end{figure}

In contrast, low-mass clouds are only weakly gravitationally bound with masses of $2.2\times 10^4\msolar$ (corresponding to $9~$per~cent of their Bonnor-Ebert mass). They have a higher central temperature yielding a lower central density with respect to thermal equilibrium (Fig.~\ref{fig:both-temp_diss}). Consequently, the central pressure is low and a smaller radius is obtained. Low-mass clouds reside in a \him{} with $T_{\rm\him}=5.5\times 10^6~$K and $n_{\rm\him}=10^{-3}\ccm$. The reference model L takes into consideration the same physical processes as model M. It is nearly isothermal, homogeneous, and isobaric with $Z=0.1\zsolar$, which leads to a radius of $28~$pc. The central temperature reads $1,160~$K and smoothly increases to $\sim 1,340~$K at the boundary. Therefore, no molecular phase is present in the cloud as dissociation is complete even in the centre (Fig.~\ref{fig:both-temp_diss}, upper plot). Model L\_nc is simulated equivalently to L, but without thermal conduction. Hence, the effect of an additional heat input on low-mass clouds can be studied.

\section{Discussion}\label{sec:discussion}

\subsection{Mass evolution}\label{subsec:massevolution}
The mass-transfer rates, $\dot{M}$, are calculated by taking the difference in mass of the cloud at two subsequent time steps, $i$ and $i+1$. So,
\begin{equation}
\dot{M}_{i\to i+1}=\frac{\mcloud(t_{i+1})-\mcloud(t_i)}{\Delta t},\label{equ:models-3}
\end{equation}
with $\Delta t=1~$Myr. The mean mass-transfer rate is given by
\begin{equation}
\langle\dot{M}\rangle=\frac 1T\sum\limits_{i=0}^T\dot{M}(\tau_{\rm sc}+i\Delta t),\label{equ:models-1}
\end{equation}
where $T=(100~{\rm Myr}-\tau_{\rm sc}[{\rm Myr}])/\Delta t$. Thus, $\langle\dot{M}\rangle$ is based on the sound-crossing time
\begin{equation}
\tsc=\frac{\rcloud}{\csbar},\label{equ:models-4}
\end{equation}
specific to each individual cloud. It is the typical time scale for a cloud to oscillate adiabatically (due to, e.g., local pressure enhancements). It is approximately the time a sound wave with mean speed $\csbar$ needs to travel into the centre and is therefore a measure for how fast a pressure balance can be established by pure hydrodynamics. We neglect cloud evolution before $\tau_{\rm sc}$ in computing $\langle\dot{M}\rangle$ in order to exclude dynamical superimposing of repulsion resulting from initial onset of mass transfer. The cloud mass at a particular time, $t_i$, is given by
\begin{equation}
\mcloud(t_i)=\sum\limits_{j=1}^N\varrho_j(t_i)V_j(t_i)=V\sum\limits_{j=1}^N\varrho_j(t_i),\label{equ:models-2}
\end{equation}
i.e. the direct sum is taken over all grid cells $j$ that are identified to belong to the cloud with constant volume $V$ of a grid cell. A cell $j$ is part of the cloud if both $\varrho_j>100\varrho_{\rm amb}$ and $T_j<5\times 10^6~$K. The values for $\dot M$ calculated by equation~(\ref{equ:models-3}) coincide with the values obtained by $\dot M=4\pi r^2\varrho(r)v(r)$ at large distances $r$ from cloud. 

\begin{figure}
\centering
\includegraphics[width=\linewidth]{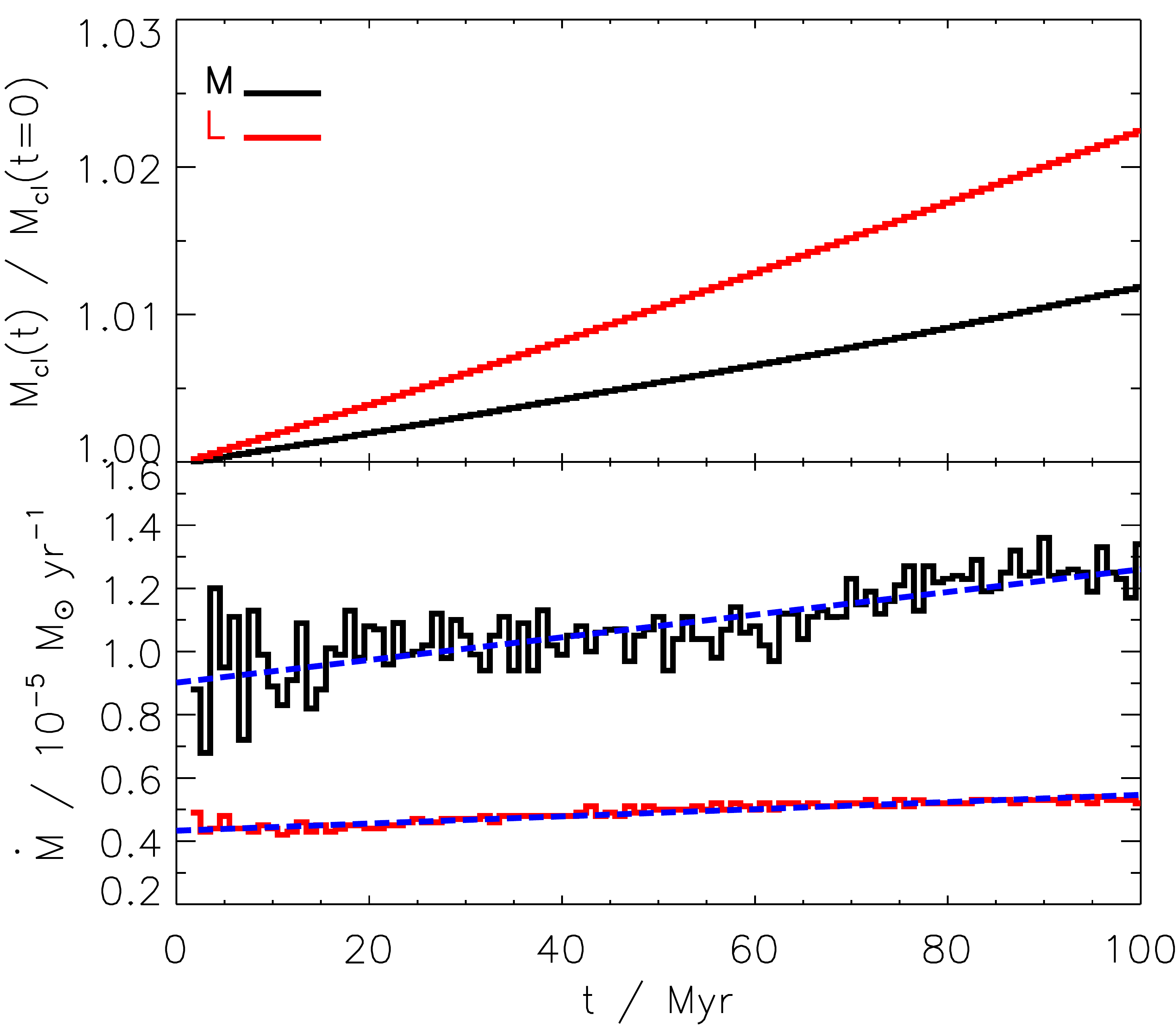}
\caption{Mass evolution for all clouds with thermal conduction. \emph{Upper plot:} Evolution of cloud masses normalized to the respective initial cloud mass. \emph{Lower plot:} Rates of condensation of ambient gas onto the clouds ($\dot M>0$) with respective linear regression (\emph{blue dashed lines}).}
\label{fig:both-mdot}
\end{figure}

It is evident from Fig.~\ref{fig:both-mdot} that ambient gas condenses onto clouds with thermal conduction. In general, the process of condensation applies to massive and low-mass clouds. By this, the initial mass of the massive (low-mass) cloud is increased by $\approx 1~$per~cent ($\approx 2~$per~cent) after $100~$Myr of evolution. The condensation rates increase over time specific to cloud mass: $3.6\times 10^{-8}~$M$_\odot~$yr$^{-1}$~/~Myr (massive cloud) and $1.1\times 10^{-8}~$M$_\odot~$yr$^{-1}$~/~Myr (low-mass cloud) as can be seen in Fig.~\ref{fig:both-mdot} (lower plot).
The mean condensation rate for the massive cloud is more than twice the rate observed at the low-mass cloud at each and every time. However, the mean mass flux
\begin{equation}
\langle\Phi\rangle=\frac{1}{100-\tsc}\sum\limits_{i=\tsc}^{100}\frac{{\rm d}\mcloud/{\rm d}t}{4\pi R_{{\rm cl},i}^2}
\end{equation}
is higher in model L as shown in Fig.~\ref{fig:both-massflow}. In models M and L the mean mass fluxes of condensing \him{} are directed towards the clouds (i.e. $\langle\Phi\rangle>0$). Since mass diffusion depends on density gradient, we study the density gradient between cloud centre and boundary, i.e.
\begin{equation}
\left\langle\nabla_\rho\right\rangle =\frac{1}{100-\tsc}\sum\limits_{i=\tsc}^{100}\frac{\rho_i(\rcloud)-\rho_i(r=0)}{R_{{\rm cl},i}}.
\end{equation}
We extend the definition of $\langle\nabla_\rho\rangle$ until the cloud centre, because even in the massive cloud M there is a tiny amount of \him{} mixing into the central region (cf. Section~\ref{subsec:mixing}). From Fig.~\ref{fig:both-massflow} we learn that a steeper radial density gradient in the cloud corresponds to a lower $\langle\Phi\rangle$.

Both $\langle\nabla_\rho\rangle$ and $\langle\Phi\rangle$ are the respective average values taken between $\tsc$ and $100~$Myr of evolution.
\begin{table}
\caption{\emph{Row 1:} Mean mass-transfer rate for each cloud as being computed between its specific $\tsc$ (cf. equation~(\ref{equ:models-4})) and $100~$Myr. For comparison, the analytical rates $\dot{M}_{\rm CM}$ of CM77 (equation~(\ref{equ:satcond-10})) and $\dot{M}_{\rm DB}$ \citep[][equation~(\ref{equ:satcond-13})]{93daltonbalbus} are provided in \emph{rows 2} and \emph{3}, respectively. Positive (negative) mass-transfer rates correspond to condensation (evaporation). \emph{Row 4} contains the mean mass flux (mean mass-transfer rate per unit surface-area).Model clouds M2 and L2 only differ in metallicity from models M and L, respectively, which in turn implies a difference in radius. Thermal conduction is present in all four model clouds.}           
\label{table:models-2}
\centering          
\begin{tabular}{l c c c c}
\hline
\multirow{2}{*}{mean mass-transfer rate} & \multicolumn{4}{c}{Model} \\
 & M & M2 & L & L2 \\
\hline
$\langle\dot{M}\rangle$ [$10^{-6}~$M$_\odot~$yr$^{-1}$] & $11.1$ & $8.8$ & $4.9$ & $1.5$ \\
$\dot{M}_{\rm CM}$ [$10^{-4}~$M$_\odot~$yr$^{-1}$] & $-10.3$ & $-7.2$ & $-4.9$ & $-0.8$ \\
$\dot{M}_{\rm DB}$ [$10^{-4}~$M$_\odot~$yr$^{-1}$] & $-11.1$ & $-7.8$ & $-5.3$ & $-0.9$ \\
\hline
$\langle\Phi\rangle$ [$10^{-10}~$M$_\odot~$yr$^{-1}~$pc$^{-2}$] & $3.2$ & $3.7$ & $4.4$ & $4.8$ \\
\hline
\end{tabular}
\end{table}
To quantify the $\langle\nabla_\rho\rangle$-$\langle\Phi\rangle$ relation we add two further model clouds: M2 has a mean radius {of $41~$pc lying between models M and L and L2 has a mean radius of $11~$pc, which is smaller than the radius of model L (M2 and L2 are identical to models M1\_me and L1\_me with solar metallicity in SH21). The mean mass fluxes result in $\langle\Phi\rangle=3.7\times 10^{-10}~$M$_\odot~$yr$^{-1}~$pc$^{-2}$ (model M2) and $\langle\Phi\rangle=4.8\times 10^{-10}~$M$_\odot~$yr$^{-1}~$pc$^{-2}$ (model L2). In order to be consistent with the models in SH21 the cooling efficiency of the plasma depends on $Z$. Metal lines dominate the line cooling for $T\gtrsim 2\times 10^4~$K \citep[][]{89boehringerhensler}. Because the cloud boundary is at $T\lesssim 10^4~$K an enhanced metallicity is negligible for determining $\langle\Phi\rangle$. All four models are well approximated by a linear $\langle\nabla_\rho\rangle$-$\langle\Phi\rangle$ relation with a slope of $51~$pc$^2~$yr$^{-1}$.

Density gradient $\langle\nabla_\rho\rangle$ and mass flux $\langle\Phi\rangle$ strongly correlate inversely with a Pearson's R-value of $-0.977$. The correlation is significant on a level of $0.05$ (p-value of $0.023$). We conclude that homogeneity of clouds leads to an enhanced mass flow of ambient material onto clouds and this mass flow is due to thermal conduction only. The standard deviation for the smallest model cloud L2 is very high (Fig.~\ref{fig:both-massflow}), because its radius changes a lot during its evolution. So, the results for the nearly homogeneous clouds cannot be interpreted uniquely and we cannot propose for sure that $\langle\Phi\rangle$ converges to a constant value. Here we emphasize, that $\langle\Phi\rangle$ is due to thermal conduction only. If thermal conduction is not accounted for, there is no mass flow towards a hydrostatic cloud in thermal equilibrium including plasma cooling (cf. Section~\ref{subsec:mixing}). Analytical mass-transfer rates predict evaporation for all clouds (Table~\ref{table:models-2}) and are thus not appropriate to model time-dependent conditions for thermal conduction.
\begin{figure}
\centering
\includegraphics[width=\linewidth]{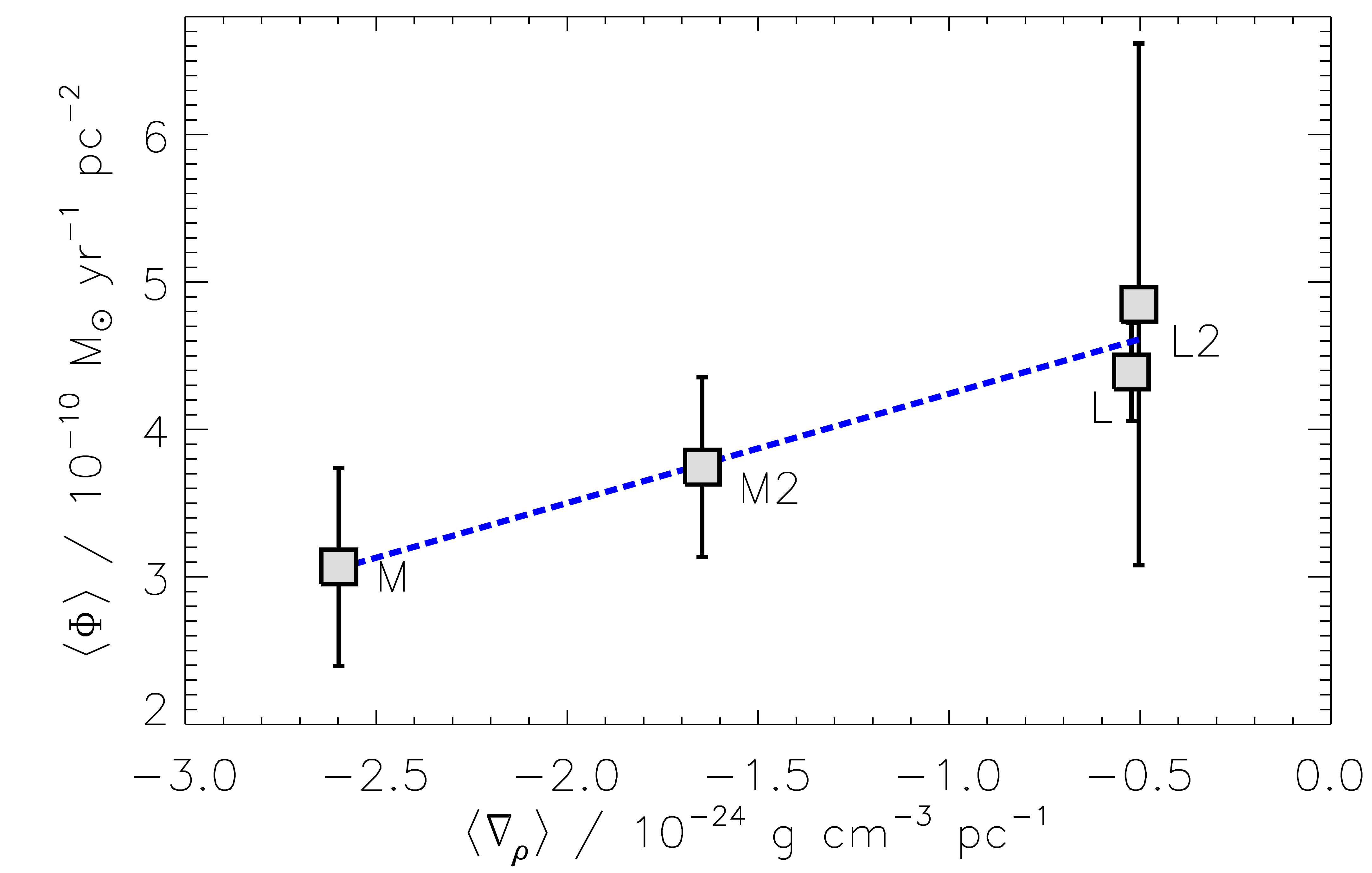}
\caption{Mean mass flux due to thermal conduction onto clouds with respect to mean radial density gradient ($\langle\Phi\rangle>0$ means condensation). The \emph{blue dashed line} is the linear regression with a slope of $51~$pc$^2~$yr$^{-1}$. The errorbars give the standard deviation for each model cloud.}
\label{fig:both-massflow}
\end{figure}

\subsection{Mixing with ambient gas}\label{subsec:mixing}
According to the continuous mass gain for the heat conducting clouds M and L we suppose the velocity field of ambient gas to be directed towards the clouds over the entire evolution indicating condensation of ambient gas. In Fig.~\ref{fig:both-density} condensation is illustrated with maximum gas velocities close to $43\kms$. As expected, in both models without thermal conduction there is only a tiny propagating net inflow of ambient \him{} due to gravity. The inflow velocity does not exceed $0.1\kms$. So, their density structures are not changed at all irrespective of cloud mass. The result is not surprising as it proves the correct numerical treatment of hydrodynamics in an equilibrium setup. Thus, only clouds with thermal conduction condense material at a reasonable rate (Table~\ref{table:models-2}). The central density peak in model M is smeared out by time and the cloud homogenizes before $50~$Myr hence affected by thermal conduction. By this, the central cooling efficiency is reduced.
\begin{figure}
\centering
\includegraphics[width=\linewidth]{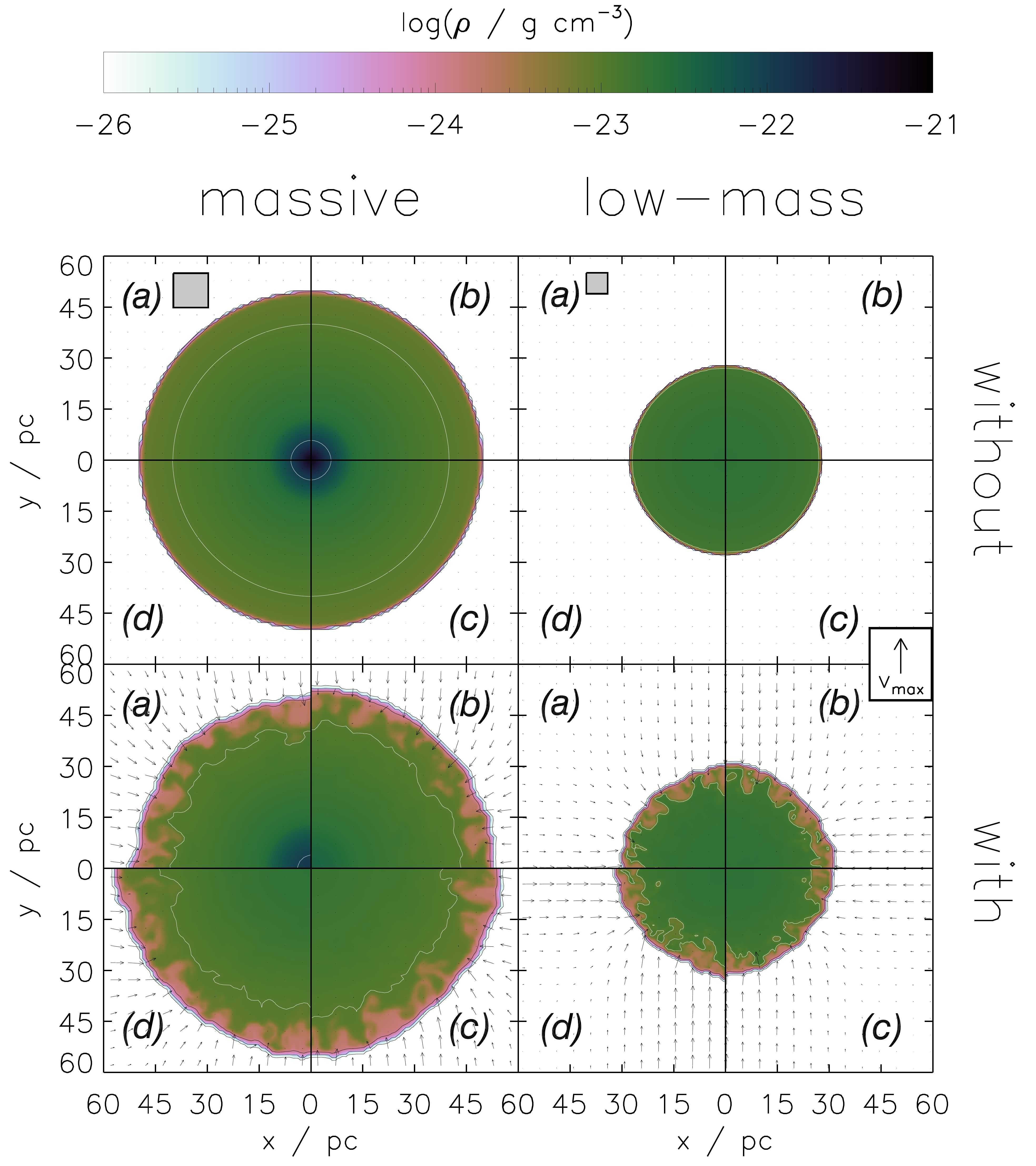}
\caption{Evolution of density in the model clouds. The plots show the section plane $(x,y)$ through cloud centre. The snapshots are taken after \snapone{} \emph{(a)}, \snaptwo{} \emph{(b)}, \snapthree{} \emph{(c)}, and \snapfour{} \emph{(d)} of evolution for clouds without (\emph{upper panel}) and with (\emph{lower panel}) thermal conduction. Isodensity contours are shown for $\varrho = 10^{-25}\ldots 10^{-21}~$g~cm$^{-3}$ separated by $\Delta(\log\varrho)=1~$dex (with the inner two contours being white just for a better contrast). \emph{Black arrows} illustrate the motion of gas scaling
linearly with respect to the maximum velocity $v_{\rm max}=43\kms$ as shown in the box to the right. The \emph{grey square} in the upper left corner is spanned by $10\,\times\,10$ cells of finest numerical resolution ($\Delta x=1.0~$pc for massive clouds and $\Delta x=0.6~$pc for low-mass clouds).}
\label{fig:both-density}
\end{figure}

Condensation of ambient material and thus an efficient mixing with \him{} is only possible in clouds with thermal conduction, which coincides with the findings of VH07b. In Fig.~\ref{fig:both-mixing} the mixing of \him{} into the cloud is shown by the fraction of \him{} to cloud gas, $f_{\rm\him}$. In models M and L an outer mixing zone evolves in thickness where $f_{\rm\him}$ can easily exceed $10~$per~cent. \him{} is mixed with cloud material and is transported into deeper regions of the cloud while without thermal conduction the clouds are not enriched by \him. The filling factor of \him{} in model L is larger than in M indicating that accreted material can be distributed easier. After $100~$Myr the entire cloud M is interspersed with at least traces of \him.

We measure the time at each cloud radius, which is needed by the mixing front to enhance the \him{} fraction at this particular radius to $1~$per~cent. In Fig.~\ref{fig:both-mixingtime} it is obvious that in model M a \him{} fraction of $1~$per~cent is reached later at every radius compared to model L.
\begin{figure}
\centering
\includegraphics[width=\linewidth]{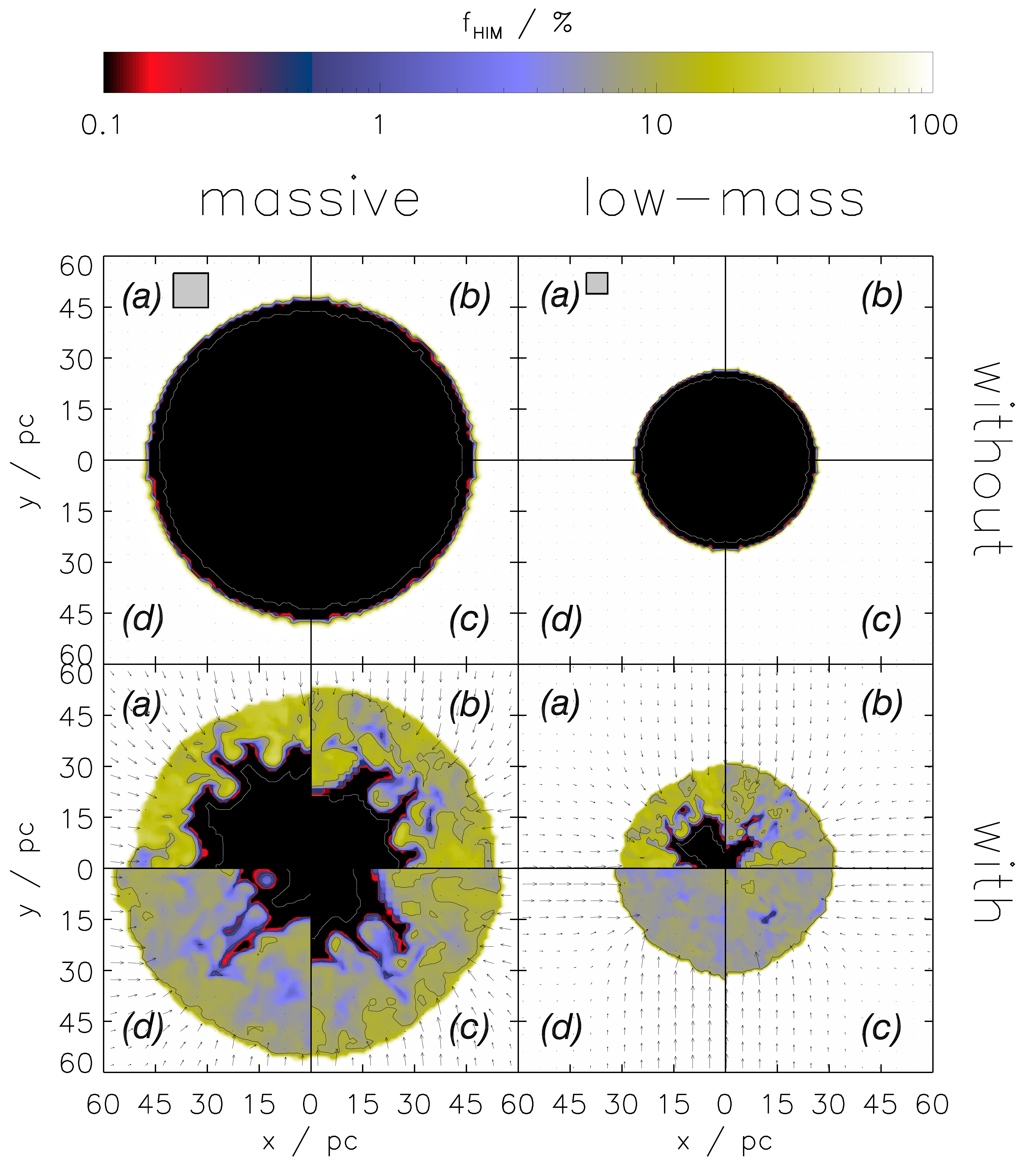}
\caption{Same as in Fig.~\ref{fig:both-density}, but for the fraction $f_{\rm\him}$ of ambient \him{} mixed into the clouds. Isocontours are shown for $f_{\rm\him} = 0.1$, $1$, and $10~$per~cent. Inside of the \emph{white contour} there is no \him{} at all.}
\label{fig:both-mixing}
\end{figure}
In model M, the mixing front only propagates until $0.3\rcloud$ (corresponding to a radius of $15~$pc) within $100~$Myr. At exactly this radius the initial central density core profile flattens (see fig.~2 in SH21). We emphasize that in model M the central cloud region will be mixed with \him{} on a much longer time scale than $100~$Myr. Even though the steep density gradient gets shallower by time (cf. Fig.~\ref{fig:both-density}) the propagation velocity of the mixing front is substantially lowered. In contrast, in the low-mass cloud, which has an almost flat radial density profile, the mixing front approaches the cloud centre within $65~$Myr. Around $0.28\rcloud$ ($\approx 8~$pc) the mixing front in model L decelerates and its velocity remains on a lower level inside this central cloud region. By taking the inverse of the piecewise slopes we can estimate the propagation velocity in model L from Fig.~\ref{fig:both-mixingtime}: for the outer $20~$pc (from $\rcloud$ to $r=8~$pc) the mixing front needs $30~$Myr yielding $m_1^{-1}\approx 0.67~$pc Myr$^{-1}$. Likewise, $m_2^{-1}\approx 0.35~$pc Myr$^{-1}$ for the central region inside $r=8~$pc in model L.
\begin{figure}
\centering
\includegraphics[width=\linewidth]{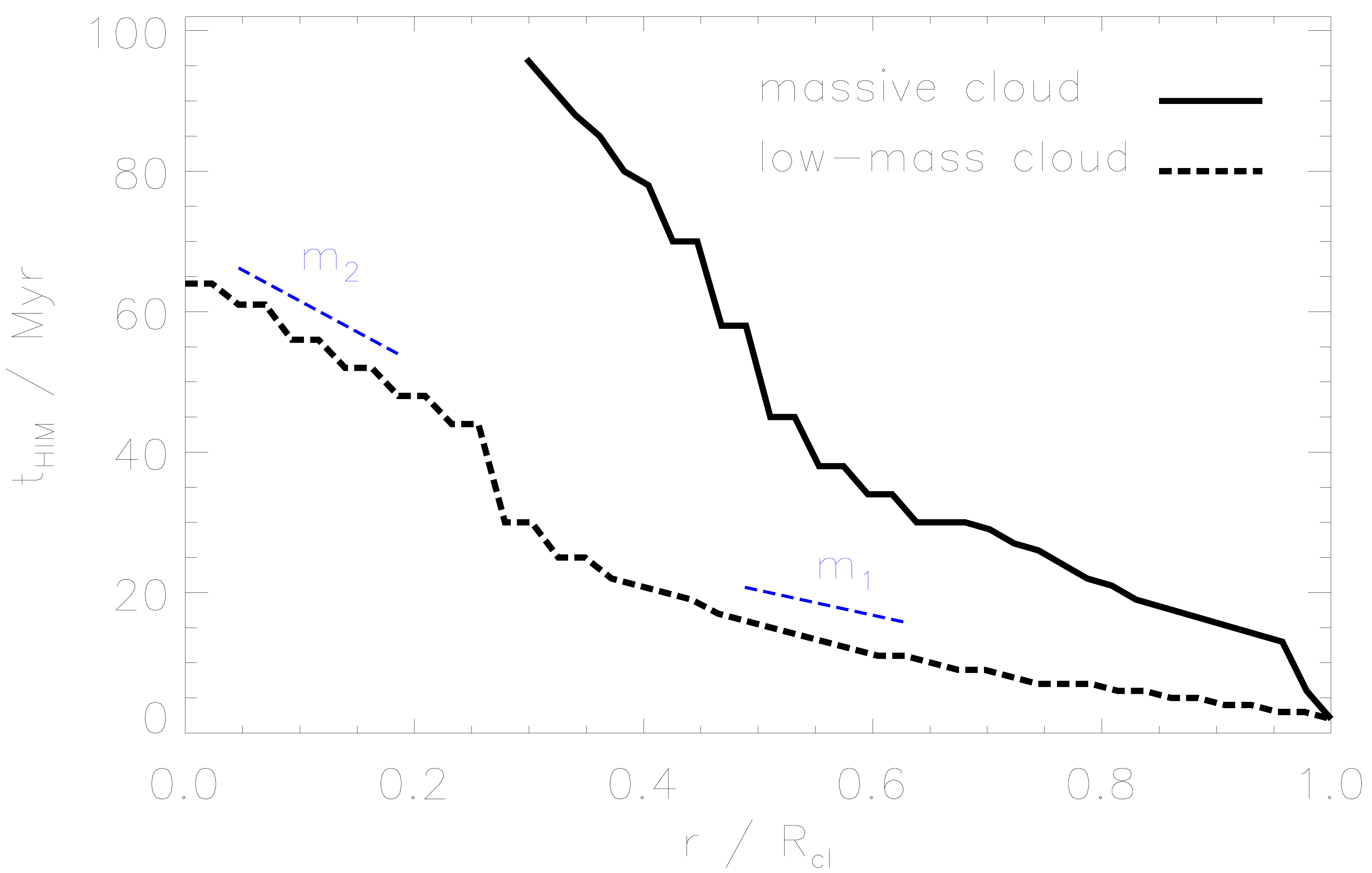}
\caption{Time after which fraction of condensed \him{} reaches $1~$per~cent in models M (\emph{solid line}) and L (\emph{dashed line}) at a given cloud-centric distance. The \emph{short blue lines} represent the piecewise slopes $m_1=-1.5~$Myr~pc$^{-1}$ and $m_2=-2.9~$Myr~pc$^{-1}$ for the dashed line. The radii are normalized to the respective cloud radius.}
\label{fig:both-mixingtime}
\end{figure}

In order to analyze the change in propagation velocity of mixing fronts in models L at $8~$pc and M at $15~$pc we show in Fig.~\ref{fig:both-mixingspectrum} (upper plot) the average \him{} fractions in the respective inner and outer region in both clouds. As expected, in model M there is almost no \him{} inside of $15~$pc while radially outwards the fraction is roughly constant after $20~$Myr with $f_{\rm\him}\approx 3~$per~cent. In model L the mixing evolves distinctly. Accreted \him{} is quickly transported into the inner region, because there is no density gradient retarding the mixing front. After $55~$Myr the mean of $f_{\rm\him}$ inside a radius of $8~$pc exceeds the \him{} fraction more outside. \him{} accumulates centrally in cloud L, but does not turn back to the outer regions due to gravity. Both dashed lines in the upper plot in Fig.~\ref{fig:both-mixingspectrum} may converge after $100~$Myr. We thus conclude that the density profile of a cloud not only affects the spatial distribution of accreted gas, but also the direction of mixing: in model M the radial gradient is sufficiently steep to provide a reasonable obstacle to slow down propagation of accreted gas and the cloud is filled up with \him{} from the outside to the inside. In contrast, if the cloud is nearly homogeneous, condensed \him{} straightly gets through into the centre, accumulates there and the cloud fills up in the reverse direction.

It is readily apparent from Fig.~\ref{fig:both-mixingspectrum} (lower plot) that only $50~$per~cent of cloud volume in model M contain more than $1~$per~cent of \him{} after $100~$Myr, while this fraction reaches $80~$per~cent in cloud L. However, the cloud regions with highest \him{} concentrations are observed in model M: within $0.7~$per~cent of the total cloud volume the \him{} fraction exceeds $10~$per~cent. We conclude again that the radial density gradient in model M accumulates in higher concentrations in the outer regions. Consequently, condensed \him{} is much easier distributed in homogeneous clouds. Hence, the above result supports the findings in \citet[][]{19sanderhensler}, that self-gravity crucially affects the evolution of even sub-Jeans clouds.
\begin{figure}
\centering
\includegraphics[width=\linewidth]{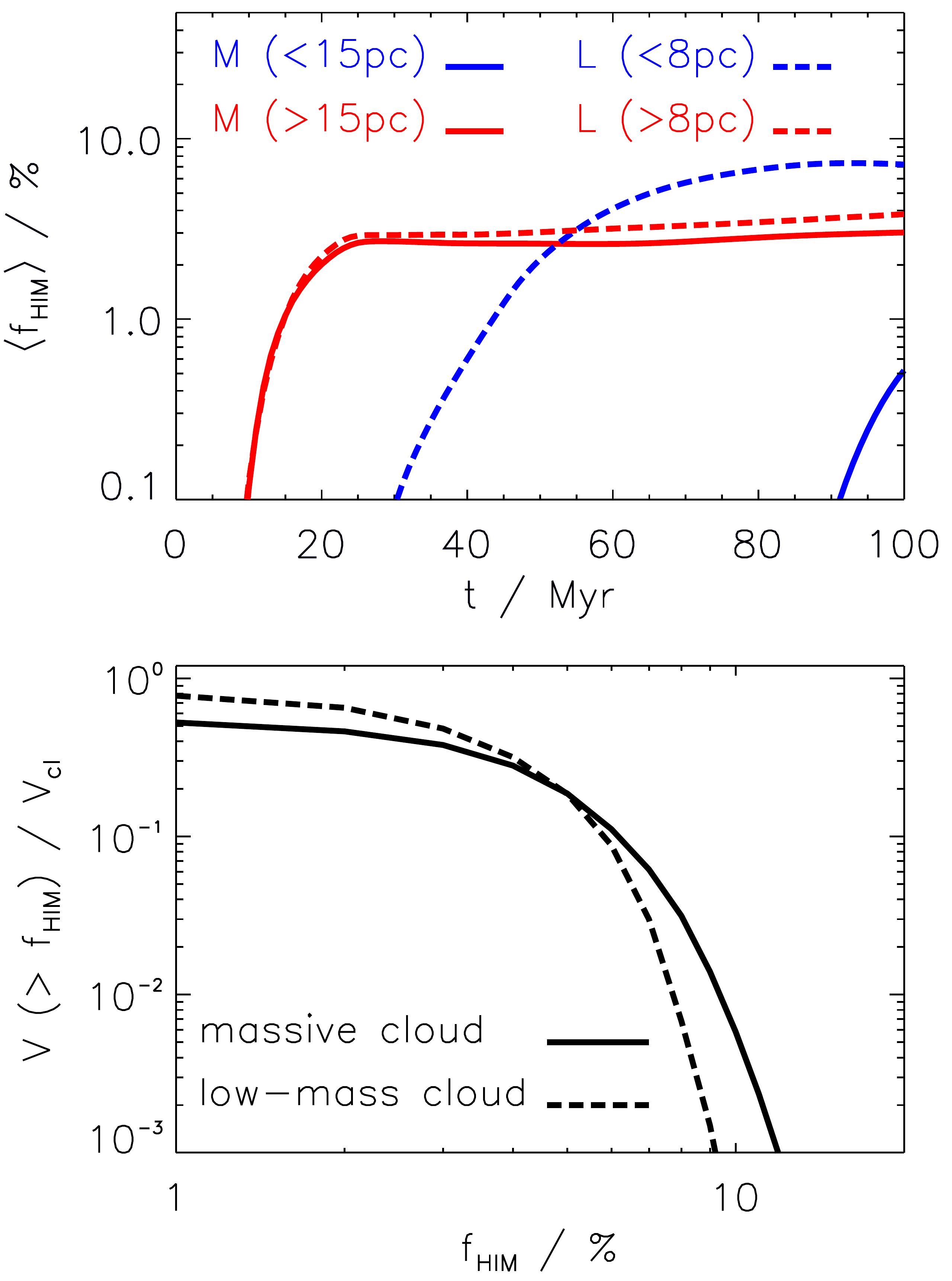}
\caption{\emph{Upper plot:} Evolution of mean \him{} fraction in model M inside and outside a radius of $15~$pc (\emph{solid lines}) and in model L inside and outside a radius of $8~$pc (\emph{dashed lines}). \emph{Lower plot:} Normalized cumulative distribution of total cloud volume containing mixed \him{} above a certain fraction after $100~$Myr of evolution.}
\label{fig:both-mixingspectrum}
\end{figure}
Mixing in moving \hvcs{} without thermal conduction is analyzed by \citet[][]{14grittonsheltonkwak}. In their 3D simulations they observe fractions of halo gas in their model cloud B between $1$ and $10~$per~cent for different ionization stages of oxygen and carbon after $100~$Myr (see their figures 3 and 4).

A simple way to trace the \him{} fraction that condenses onto the clouds is realized by studying the mean metallicity in each particular grid cell. The mean metallicity in both models M and L increases on average, while it is more enhanced in the low-mass cloud (upper plot in Fig.~\ref{fig:both-metltime}). The growth rates due to condensation of ambient gas are roughly $0.012\zsolar/100~$Myr (model M) and $0.014\zsolar/100~$Myr (model L) and are rather low. The fact that the clouds condense \him{} at a rather low rate coincides with the findings of \citet[][]{21deciaetal} that low-metallicity gas falling through the hot halo of the Milky Way towards the Galactic disk as \hvcs{} sustains observed chemical inhomogeneities on scales of $10\ldots 100~$pc. The \hvcs{} thus would be different in metallicity for reasonable time scales (at least $100~$Myr), because the cloud gas does not mix effectively with surrounding halo gas.
\begin{figure}
\centering
\includegraphics[width=\linewidth]{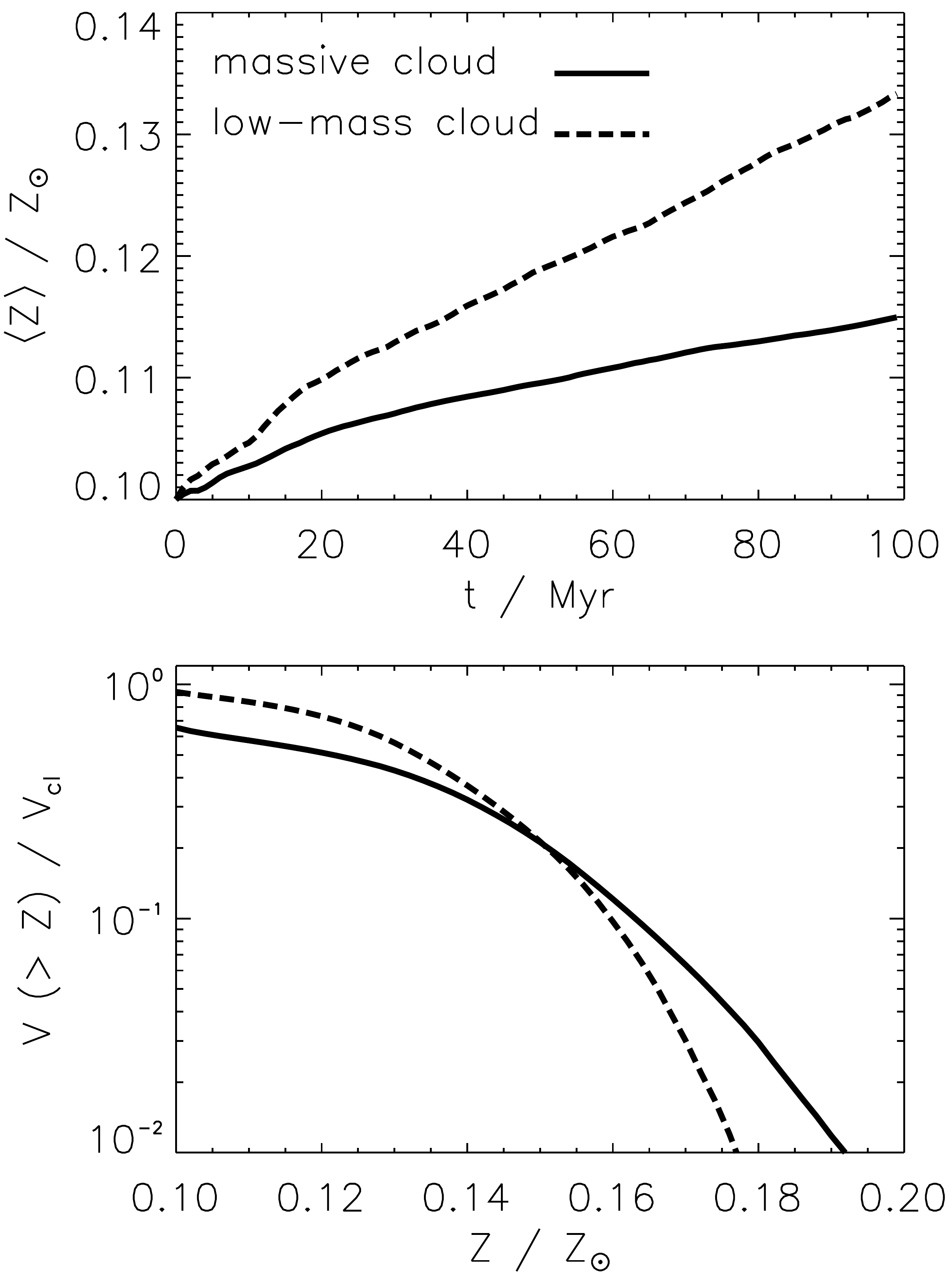}
\caption{\emph{Upper plot:} Evolution of density-weighted metallicity in model M within a radius of $40~$pc (\emph{solid line}) and in L within a radius of $20~$pc (\emph{dashed line}). \emph{Lower plot:} Normalized cumulative distributions of metallicity in both model clouds after $100~$Myr of evolution.}
\label{fig:both-metltime}
\end{figure}

By inspecting the lower plot in Fig.~\ref{fig:both-metltime} one finds that the regions of highest metallicity are found in the massive model cloud. The dynamics of metals is consistent with the mixing of condensed \him{} (see discussion above) hence supporting the reasoning, that it is easier for accreted gas to spread in a cloud with only a shallow radial density gradient. Vice versa, if a substantial density gradient is present in the central region then the accreted material accumulates preferentially in the outskirts. The metallicity in the outer regions in stratified clouds can exceed the metallicities found in homogeneous clouds. The distribution of \him{} in the model clouds traces their enrichment by heavy elements if a metal contrast is given between cloud and \him.

\subsection{Structure of interface}\label{subsec:interface}
During the evolution of both models M and L a transition zone forms at the interface between cloud and ambient, hot gas. From the upper plot in Fig.~\ref{fig:both-equality} it is obvious, that the condensation velocity of ambient gas is negative over the entire evolution thus condensation is continuous. The condensation velocity is between $-20\ldots -10\kms$ and is thus much smaller than the sound speed in the hot ambient gas and nearly one order of magnitude smaller than the local sound speed. But within the transition zone the condensation velocity is remarkably lowered by one order of magnitude to $\lesssim 2\kms$. This velocity is distinctly smaller than the sound speed in the local warm cloud gas. By this, material is accreted faster than being distributed within the clouds. The mean velocity is lowered, because the temperature in the clouds is much lower at a few $10^3~$K. The inflow of hot ambient gas enhances the particle density thus increasing the cooling efficiency. By this, transfer of momentum from hot conducted electrons to cloud gas is compensated by an enhanced cooling.  So, gas heating by conduction is not sufficient to overcome cooling in the outskirts of the clouds. The mixing of \him{} with the cloud gas in deeper layers needs longer than the accretion of \him: while \him{} is accreted with a velocity of $\sim 12~$per~cent of local sound speed it is transported into deeper cloud regions with a local Mach number of $0.03$ only. \him{} is thus both condensed and distributed within the clouds only subsonically (according to the respective local sound speed).
\begin{figure}
\centering
\includegraphics[width=\linewidth]{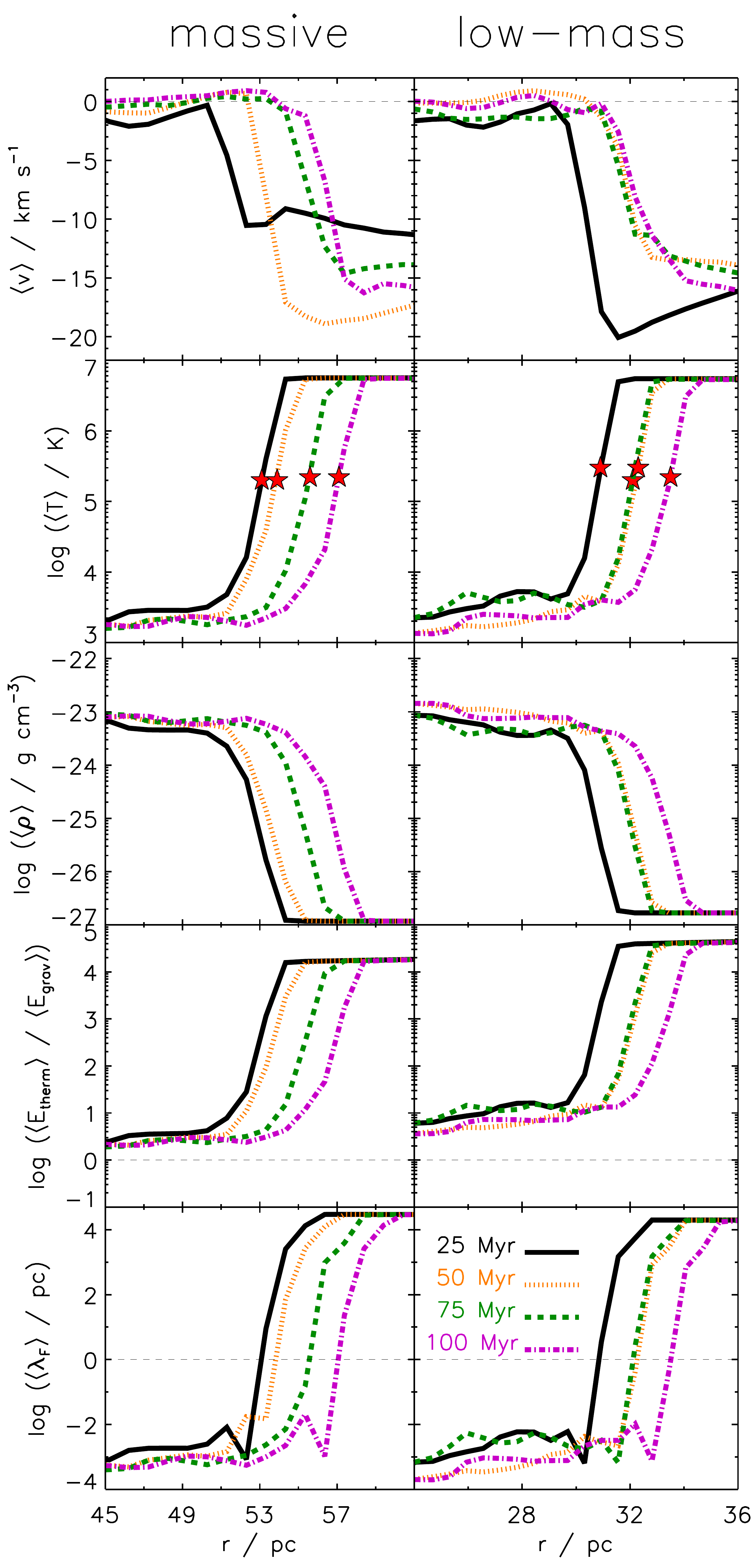}
\caption{\emph{From top to bottom:} Evolution of angle-averaged, radial profiles for inflow velocity, temperature, density, ratio of thermal to gravitational energy, and Field length around the boundaries of clouds with thermal conduction shown at four diffent snapshots. The \emph{red stars} denote the radii at which $\lf=1$, i.e. the strength of thermal conduction equals that of radiative cooling.}
\label{fig:both-equality}
\end{figure}

Within the transition zone the temperature raises substantially from a few $10^3~$K until it reaches $\tamb$. As the cooling function scales with $\varrho^2$ and sharply increases towards $T\approx 10^4~$K the cooling efficiency has a maximum within the transition zone: for radii smaller than the inner boundary of the transition zone the temperature is too low. Radially outward of the transition zone the density is too low. Thus balance between maximum of cooling and energy input by thermal conduction is reached within the transition zone, i.e. here $\lf =1$, which is shown by red stars in Fig.~\ref{fig:both-equality}. At $\lf =1$ an equilibrium temperature of $2\ldots 3\times 10^5~$K is reached. Temperatures around $10^4~$K can only be reached if the density is about $100$ times higher, which is the case at smaller cloud-centric distances and can be seen by inspecting the density profile in Fig.~\ref{fig:both-equality}. Outside of the transition zone $\lf$ reaches values of $\gtrsim 5\times 10^2\rcloud$. Inside of the transition zone the cooling remains on a reasonable level thus $\lf$ immediately drops below $10^{-2}\rcloud$ for $r<r(\lf=1)$ and thermal energy is lowered such that $E_{\rm therm}\gtrsim E_{\rm grav}$. By this, hot ambient gas condenses onto the clouds while theoretical mass-loss rates predict evaporation (Table~\ref{table:models-2}).
\begin{figure}
\centering
\includegraphics[width=\linewidth]{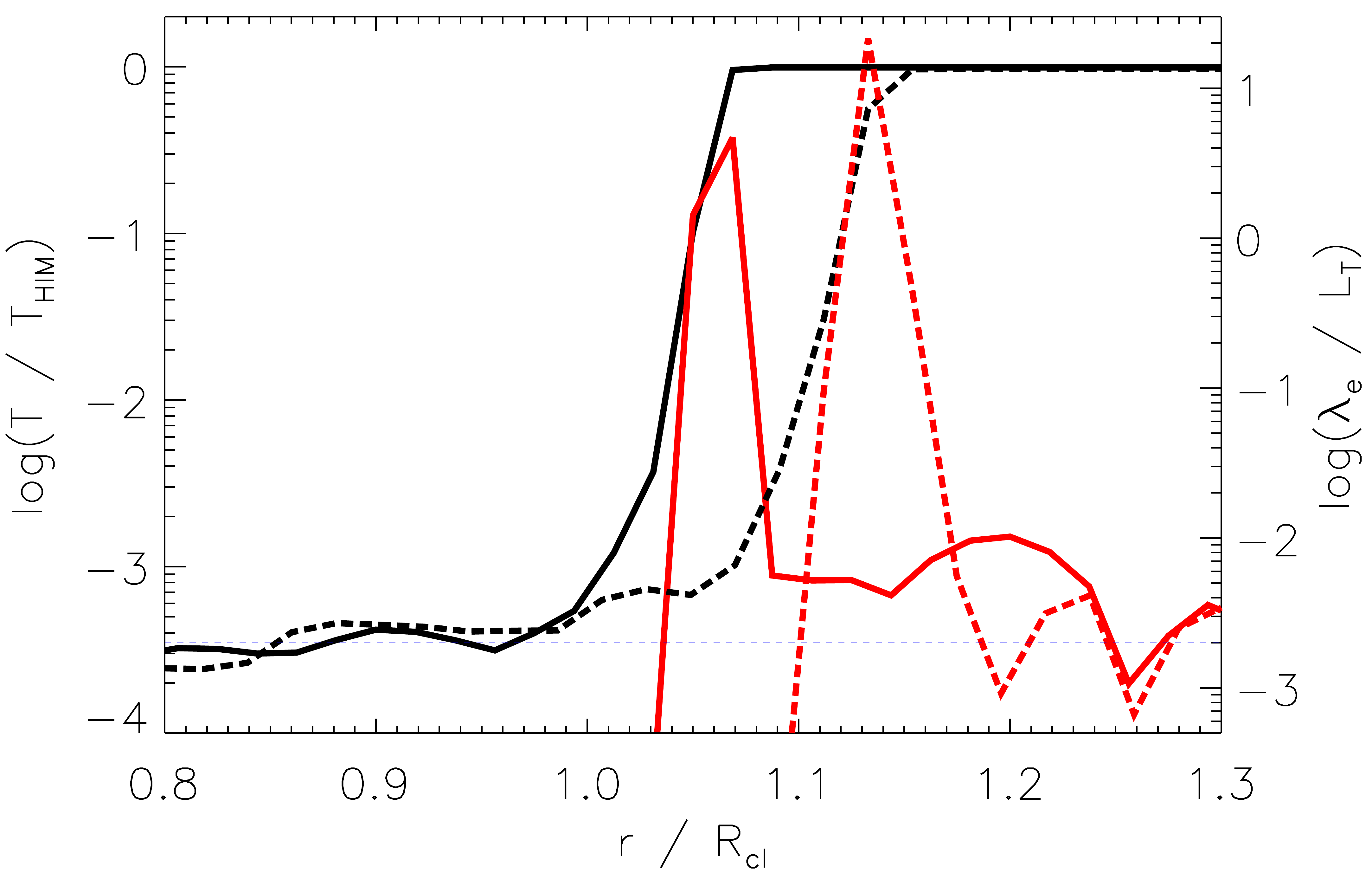}
\caption{Temperature profile normalized to ambient temperature over inner classical, saturation, and outer classical zone for models M (\emph{black solid line}) and L (\emph{black dashed line}) after $100~$Myr of evolution. The ratio $\lel/\LT$ (\emph{red line} for respective model cloud) distinguishes between classical and saturated thermal conduction (threshold $\lel/\LT = 2\times 10^{-3}$ (equation~(\ref{equ:satcond-2})) is indicated by the \emph{thin dashed line)}.}
\label{fig:both-temp_zones}
\end{figure}

We compare our findings to the theory described in CM77. They infer from the enthalpy flux that the interface between cloud and \him{} is divided into three zones: (1) an \emph{inner classical zone} extending from $\rcloud$ outward to a distance $R_{\rm sat,1}$, where the heat flux becomes saturated. The heat flux in the inner zone is classical, because the temperature and hence the conductivity are small. (2) A \emph{saturation zone} extends between $R_{\rm sat,1}$ and $R_{\rm sat,2}$. The temperature increases steeply, such that the temperature gradient is large. Thus, the heat flux is saturated. (3) Outside of $R_{\rm sat,2}$, in the so-called \emph{outer classical zone}, the temperature gradient is small, because the temperature level of the \him{} is reached. Consequently, the heat flux is classical again. At these high temperatures the cooling function dominates the input of conducted thermal energy (equilibrium is marked by red stars in Fig.~\ref{fig:both-equality}).

CM77 analytically deduce the values for $R_{\rm sat,1}$ and $R_{\rm sat,2}$ based on both the Mach number $M_s$ in the saturated zone and $\sigma_0$ (see their equations $48$ and $58$, respectively). The values for $M_s$ and $\sigma_0$ do not change in the analytically calculated saturation zone, because the temperature profile is constant in CM77.
\begin{table}
\caption{Radial extension of inner classical zone (ranging from $\rcloud$ to $R_{\rm sat,1}$), saturation zone (ranging from $R_{\rm sat,1}$ to $R_{\rm sat,2}$), and outer classical zone (above $R_{\rm sat,2}$) according to CM77 (columns labeled `CM77') and based on simulation results (columns labeled `Sim.'). The Mach numbers are averaged over inner classical zone (icz) or saturation zone (sz) while the ratios $\rcloud/\lf$ are taken at particular zone transitions. All values are time averaged over $100~$Myr. The uncertainties are standard deviations.}
\label{table:theoryCM77}
\centering          
\begin{tabular}{l c c c c}
\hline
mean & \multicolumn{2}{c}{model M} & \multicolumn{2}{c}{model L} \\
values & CM77 & Sim. & CM77 & Sim. \\
\hline
$\rcloud~$/~pc & \multicolumn{2}{c}{$52\pm 3$} & \multicolumn{2}{c}{$30\pm 2$} \\
$R_{\rm sat,1}~$/~pc & $58\pm 3$ & $54\pm 2$ & $33\pm 2$ & $31\pm 1$ \\
$R_{\rm sat,2}~$/~pc & $77\pm 3$ & $60\pm 4$ & $46\pm 3$ & $35\pm 2$ \\
$Ma$(icz) & $0.09\pm 0.05$ & $0.17\pm 0.08$ & $0.2\pm 0.1$ & $0.2\pm 0.1$ \\
$Ma$(sz) & $0.04\pm 0.01$ & $0.05\pm 0.02$ & $0.04\pm 0.01$ & $0.06\pm 0.02$ \\
$\rcloud/\lf$($\rcloud$) &  & $250$ &  & $500$ \\
$\rcloud/\lf$($R_{\rm sat,1}$) &  & $0.38$ &  & $0.26$ \\
$\rcloud/\lf$($R_{\rm sat,2}$) &  & $3\times 10^{-5}$ &  & $5\times 10^{-5}$ \\
\hline
\end{tabular}
\end{table}
Based on the temperature profile in Fig.~\ref{fig:both-temp_zones} we identify the transition zone that forms around our simulated clouds with the inner classical zone and the saturation zone from CM77.

We observe in our simulations that the saturation zone is located closer to the cloud boundary than in analytic calculations by CM77. Hence, the inner boundary of saturation zone $R_{\rm sat,1}$ is on average closer to $\rcloud$ (Table~\ref{table:theoryCM77}). Likewise, the saturation zone $\Delta R_{\rm sat}\equiv |R_{\rm sat,2}-R_{\rm sat,1}|$ is narrower in our simulations. From Table~\ref{table:theoryCM77} we deduce $\Delta R_{\rm sat}=6\pm 4~$pc in model M and $\Delta R_{\rm sat}=4\pm 2~$pc in model L, or, in terms of the respective cloud radius, $\Delta R_{\rm sat}/\rcloud=1.04\ldots 1.15$ (model M) and $\Delta R_{\rm sat}/\rcloud=1.03\ldots 1.17$ (model L). This agrees with the results of VH07b. Our model clouds are distinct from the clouds considered in CM77 in terms of plasma cooling and heating, and self-gravity.

Within the saturation zone it holds $\lel/\LT\gg 2\times 10^{-3}$ (Fig.~\ref{fig:both-temp_zones}). Hence, according to condition (\ref{equ:satcond-2}) the heat flux is saturated. By inspecting the lower plot in Fig.~\ref{fig:both-equality} we find that $\lf$ reaches values $\gtrsim 10^4~$pc inside the saturation zone. Therefore, the condition for condensation is already fulfilled for values $\rcloud\gtrsim 0.02\ldots 0.07\lf$, which decreases the threshold discussed in VH07b by a factor of at least $3$. As a consequence, condensation already sets in at larger $\lf$, which correspond to cooling rates that are one order of magnitude lower as those found in VH07b.

We conclude that the condition for condensation is met within the saturation zones in our model clouds and thus ambient hot material condenses. Our results are consistent with the condensation condition $\rcloud/\lf\gtrsim 0.24\ldots 0.36$ discussed in VH07b.

\subsection{Condensation rates}\label{subsec:condrates}

We already discussed the effects of thermal conduction, cloud mass, and radial density gradient on condensing hot ambient material onto the simulated clouds. Compared to analytic considerations of CM77 the interface structure in our simulated clouds is time-dependent and so are the condensation rates.

In Fig.~\ref{fig:transzone_thickness} we show the analytically calculated radial interface structure based on CM77. The three zones are of constant thickness over the entire evolution of the clouds. The red lines in Fig.~\ref{fig:transzone_thickness} show the respective saturation zone for our model clouds. It is visible to the eye that in model cloud L its thickness decreases by time. In model M the saturation zone gets slightly narrower (hardly noticeable in Fig.~\ref{fig:transzone_thickness}, upper plot). Simultaneously, the condensation rates $\dot M$ increase in both models (cf. lower plot in Fig.~\ref{fig:both-mdot}). Below we attempt to explain the relation between thickness of saturation zone and strength of condensation rate.

The interface between cloud and \him{} structures radially into an inner classical zone, a saturation zone, and an outer classical zone due to saturated thermal conduction (cf. CM77). In the outer classical zone it holds $\rcloud/\lf\ll 1$ (see the respective value at outer radius $R_{\rm sat,2}$ in Table~\ref{table:theoryCM77}). Within the saturation zone $\lf$ drops such that $\rcloud/\lf\sim 1$. Hence, in the inner classical zone on average $\rcloud/\lf = 0.7$ (model M) and $\rcloud/\lf = 3.1$ (model L). Consequently, the condition for condensation of gas is fullfilled within both the saturation zone and the inner classical zone. By the process of condensation, hot ambient gas is transported into saturation zone and inner classical zone at a rate $\dot M$. Therefore, the particle density $n$ grows within the saturation zone and the inner classical zone. Because the cooling efficiency is proportional to $n^2$ the cooling strength grows and hence according to equation~(\ref{equ:resolution-1}) $\lf$ decreases. But if gas in the saturation zone cools the pressure drops and $\dot M$ increases. Actually, we find an anti-correlation between $\lf$ in the saturation zone and $\dot M$ with Pearson's R-values of $-0.14$ (model M) and $-0.25$ (model L). The anti-correlation is weakly significant for model M (p-value $0.16$) and strongly significant for model L (p-value $0.01$). We further find an anti-correlation between $\dot M$ and $\Delta R_{\rm sat}$ for model M (R-value $-0.17$, weakly significant with p-value $0.09$) and model L (R-value $-0.38$, strongly significant with p-value $<0.01$). From a physical point of view, the latter anti-correlation is attributed to the increased cooling efficiency by higher $n$ due to a larger $\dot M$. A lower pressure leads to a narrower saturation zone. By this, the radial temperature gradient across saturation zone gets steeper. So, $\dot M$ increases again. We indeed find a positive correlation between the temperature gradient across the saturation zone and $\dot M$. For model M, the Pearson's R-value is $0.13$ with a weak significance on level $0.05$ (p-value $0.19$). For model L, the R-value reads $0.23$ and the correlation is strongly significant (p-value $0.02$).

The above described `condensation engine' does not stop if the reservoir of hot ambient gas is large compared to the cloud. As a result, we observe an increase of $\dot M$ and a decrease of $\Delta R_{\rm sat}$ by time.
\begin{figure}
\centering
\includegraphics[width=\linewidth]{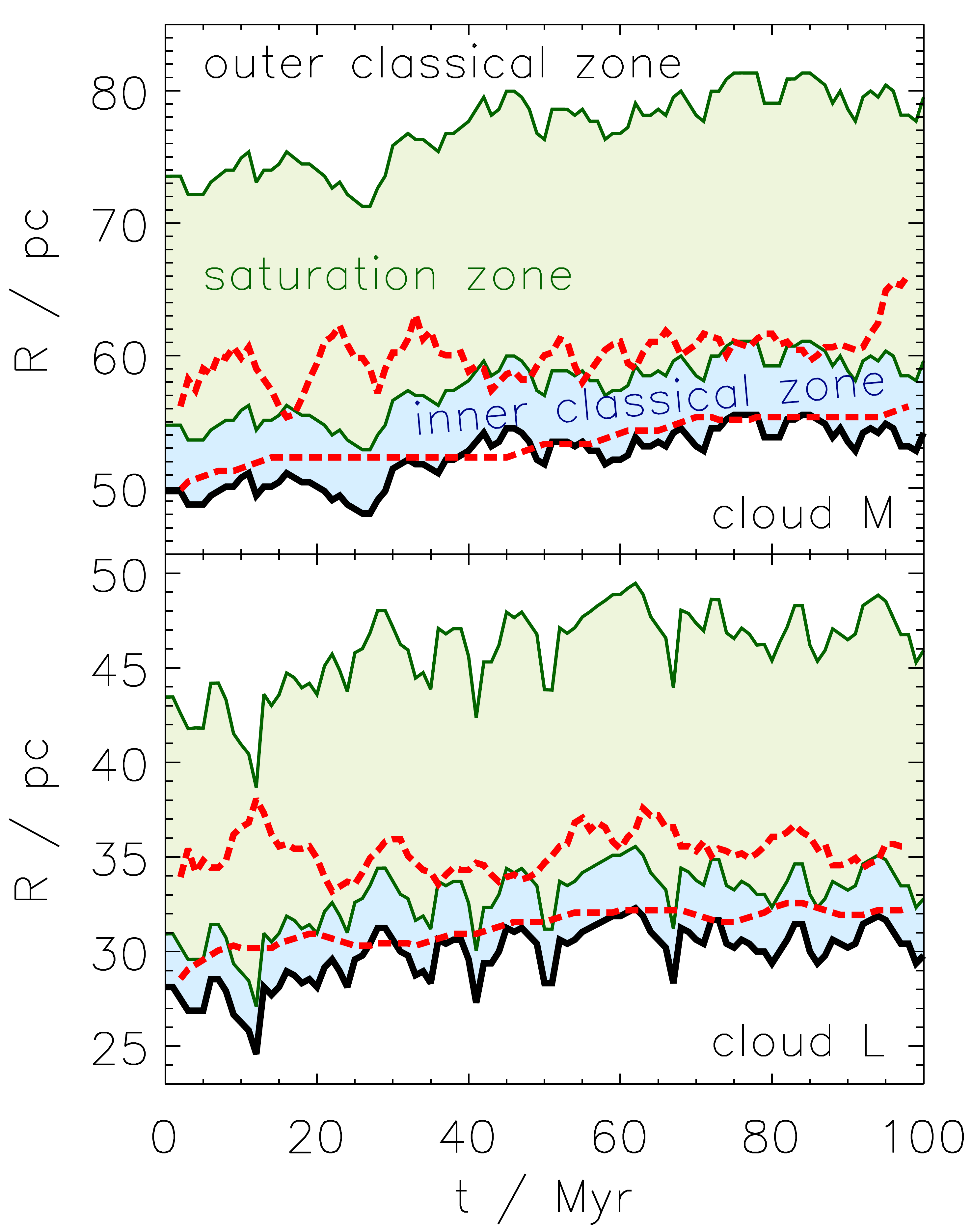}
\caption{Evolution of structure of cloud interface in massive (\emph{upper plot}) and low-mass clouds (\emph{lower plot}). The colored areas regard the analytical calculations by CM77 for inner classical (\emph{blue}), saturation (\emph{green}), and outer classical zone (\emph{white}). The \emph{red dashed lines} enclose the saturation zone of our simulations.}
\label{fig:transzone_thickness}
\end{figure}

\section{Summary and conclusions}\label{sec:sumcon}

With our simulation suite we show both the limits of the analytical approach of thermal conduction by CM77 and extend the numerical analyses performed by VH07b. We conduct three-dimensional hydrodynamics and plasma-physics simulations of interstellar multiphase clouds at rest with respect to an ambient hot, tenuous, and highly ionized gas using the AMR code {\sc Flash}. Relevant physical processes considered numerically or semi-analytically are saturated thermal conduction, self-gravity, metal-dependent heating and cooling of the plasma by radiation and collisions, dissociation of molecules and ionization of atoms, and mass diffusion. In SH21 we simulate high-velocity clouds with thermal conduction while in the present work we analyze their resting counterparts. By comparing identical model clouds, which only differ in thermal conduction, we work out differences in evolution based solely on this physical process in presence of steep temperature gradients.

While analytical mass-transfer rates by CM77 only show evaporation, \him{} condenses continuously on our model clouds during their entire evolution (Table~\ref{table:models-2}). This originates from the fact that CM77 only integrate the time-independent equation for energy conservation in the heat conducting interface. But it turns out, that the rate of condensation is closely related to the effective heat flux crossing the transition zone at cloud surface: mixing of cool cloud gas with ambient hot plasma diminishes the temperature contrast between cloud and \him{} and thus the effective heat flux reduces accordingly.

Regardless of mass, our model clouds are in the domain of condensation. This is in agreement with the two-dimensional results obtained by VH07b. Our simulated low-mass clouds are more easily heated up than the massive clouds, because their density is lower and so is the cooling rate. The lack of a substantial negative radial density gradient leads to a faster distribution of condensed ambient gas in the clouds.

Condensation is found to be a valid process to mix \him{} into the clouds. However, mixing is not equally efficient in all clouds and appears at a rather low rate. In model L condensed \him{} is distributed more easily than in model M such that cloud and ambient gas homogenize faster. The mixing velocity in model M drops at a cloud-centric distance of $\sim 15~$pc where the density gradient increases. As a consequence, \him{} accumulates outside of the steepest slope of the negative central density gradient. Because we use metallicity as a tracer of accreted \him{} we draw two conclusions: (1) if a metallicity contrast is present, then metal enrichment of clouds is likely; (2) sub-regions with highest metallicity are found in clouds with steepest radial density profiles. The main results of our study are summarized as follows:
\begin{enumerate}
\item Saturated thermal conduction is able to disturb an initial thermal equilibrium in multiphase clouds such that condensation of ambient hot gas is initiated and maintained and a continuous mixing process is triggered.
\item With thermal conduction considered, the central density peak in massive clouds levels off within $50~$Myr.
\item The velocity by which gas condenses onto clouds is higher than the mixing velocity inside the clouds. Hence, gas is accreted faster than being distributed.
\item The shallower the radial density gradient the higher the mass flux due to condensation. We conclude that homogeneity in a cloud produces a higher material flow towards it. So as not to overestimate the condensation rate, it is important to consider self-gravity even in sub-Jeans clouds, which naturally implies a radial density gradient. This supports the results in \citet[][]{19sanderhensler}.
\item Self-gravity substantially affects the spatial distribution of hot ambient material condensed onto a cloud. The noticeable radial density gradient in stratified clouds prevents condensed \him{} from being mixed into the central cloud region. Thus, \him{} accumulates in the outer regions of stratified clouds thus the highest mixing fractions are found there. If cloud and \him{} have different metal contents, the sub-regions with highest metallicity are observed in stratified clouds with a distinct radial density gradient. Consequently, a radial metallicity gradient forms. Condensed gas is more easily distributed in homogeneous clouds and hence mixing is more efficient.
\item A transition zone forms at the interface between cloud and \him{} due to saturated thermal conduction. The transition zone splits up into two adjacent zones with the same thermal characteristics described in CM77 for their inner classical zone and saturation zone. Compared to the analytic considerations in CM77 the saturation zones in our simulated clouds start at smaller radii, they are much narrower, and their thicknesses decrease by time. Within the transition zones the condition for condensation is met.
\end{enumerate}
We conclude that plasma cooling, self-gravity, and thermal conduction play a key role in the evolution of multiphase clouds with a core-halo structure in density, temperature, and pressure. Self-gravity acts as a stabilizing process against Rayleigh-Taylor instabilities, which can otherwise develop if the clouds are massive enough \citep[][]{93murrayetal}. The process of condensation provides a plausible, yet not efficient mechanism to mix ambient \him{} into the cloud. If mixing is representative for \hvcs, then their low metallicities \citep[][]{01wakker} are presumably a consequence of reduced mixing. By this, the findings of \citet[][]{21deciaetal} can be interpreted that \hvcs{} are observed to keep their chemical inhomogeneities over dozens of Myr. As our most important result we deduce that clouds can resist thermal evaporation even in those parameter ranges, where hitherto analytical approaches predict evaporation.

\section*{Acknowledgements}
This work was supported by the Doctoral College (Initiativkolleg) I033-N at the University of Vienna and by the Austrian Science Fund (FWF), project number P 21097. The software used in this work was in part developed by the DOE NNSA-ASC OASCR Flash Center at the University of Chicago. The computational results presented have been achieved using the Vienna Scientific Cluster\footnote{see \href{http://vsc.ac.at/}{http://vsc.ac.at/}}.

\section*{Data availability}

The data underlying this article (simulation results, analysis scripts) will be shared on reasonable request to the corresponding author.




\bibliographystyle{mnras}
\bibliography{references} 


%

\appendix

\section{Resolution study}\label{app:resolution}
%
%
%
%
\begin{figure*}
\centering
\includegraphics[width=.85\textwidth]{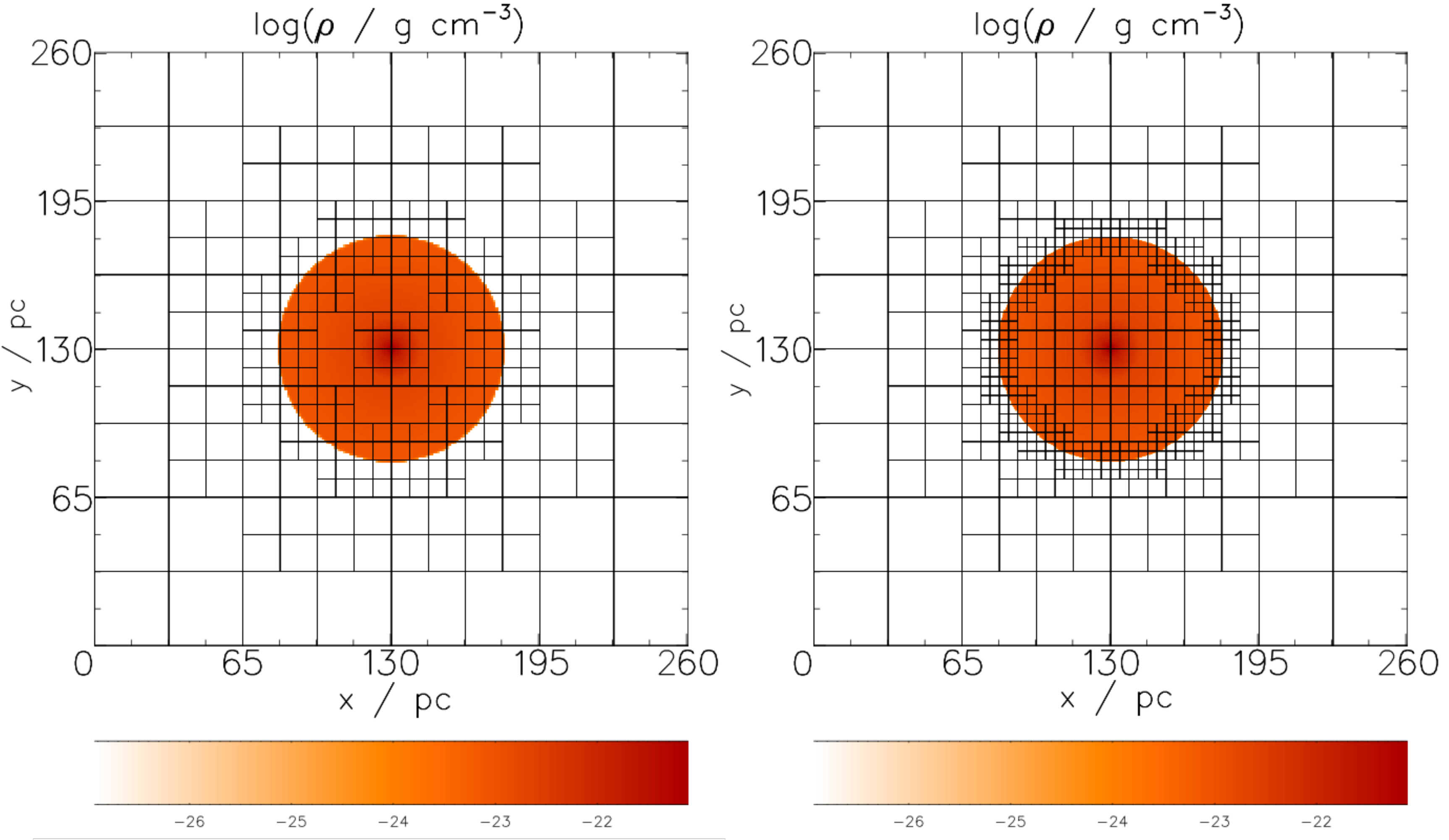}
\caption{Density slice of a massive cloud with $6$ levels of refinement ($\Delta x=1.02~$pc, \emph{left plot}) and $7$ levels of refinement ($\Delta x=0.51~$pc, \emph{right plot}). The blocks of the computational mesh are shown with one block consisting of $8\times 8\times 8$ grid cells with respective side lengths of $\Delta x$.}
\label{fig:resolution_blocks}
\end{figure*}
A comparison of the numerical to the analytical solution of pure thermal conduction in three spatial dimensions is already shown in \citet[][]{21sanderhensler}. Both solutions coincide on a reasonable level of accuracy.

In order to also proof the sufficiency of numerical resolution for the model clouds used in this work, we perform simulations with equal setup and halven the finest numerical resolution. To account for a finer grid spacing, we increase the level of refinement from $6$ (massive: $\Delta x=1.02~$pc, low-mass: $\Delta x=0.63~$pc) to $7$ (massive: $\Delta x=0.51~$pc, low-mass: $\Delta x=0.31~$pc). The mesh is adaptively refined based on the local density gradient, not on sound speed. This means, grid refinement can be coarse-grained in hot, but homogeneous regions (as is the situation far away from the cloud in the hot ambient medium). Furthermore, the mesh is conservatively refined, i.e. any two adjacent mesh regions (in terminology of the {\sc Flash} code they are called ``blocks'') are only allowed to differ in one refinement level. Therefore, cold regions close to steep density gradients at the cloud boundary and the interface itself between cloud and ambient medium are at highest possible refinement level over the entire evolution. Conclusively, this interface, where condensation takes place, is definitely resolved at level $7$ in the highly resolved models, while the internal cloud part is equally refined in both resolution setups. To illustrate the above mentioned, we show the numerical grid for the two massive model clouds in Fig.~\ref{fig:resolution_blocks}.

We compare the model clouds with low and high resolution for about $4$ sound-crossing times (cf. equation~\ref{equ:models-4}). The highly resolved models show larger variations in the magnitude of their mass-transfer rates. Consequently, they have much larger standard deviations (Fig.~\ref{fig:resolution}). Nonetheless, their mean rates $\langle\dot{M}\rangle$ do not differ by more than $2~$per~cent (cf. Table~\ref{table:resolution}). The pairwise differences in the mean mass fluxes $\langle\Phi\rangle$ are a bit higher, because they relate to the surface of respective cloud, which scales with the square of cloud radius. A finer resolution leads to a slightly smaller cloud radius. In summary,
\begin{enumerate}
\item The mean mass-transfer rates of a low resolved and a highly resolved cloud deviate by less then one standard deviation.
\item The Field length $\lf$ is already sufficiently resolved by $6$ levels of refinement.
\end{enumerate}
We hence conclude that the numerical resolution used in models M and L is sufficient to analyze processes related to thermal conduction.

\begin{table}
\caption{Model clouds M and L in comparison to their respective highly resolved (h.r.) variants. We compare the mean mass-transfer rates $\langle\dot{M}\rangle$ with standard deviations $\sigma\left(\langle\dot{M}\rangle\right)$ and the mean mass fluxes $\langle\Phi\rangle$.}
\label{table:resolution}
\centering          
\begin{tabular}{l c c c c}
\hline
\multirow{2}{*}{values} & \multicolumn{4}{c}{Model} \\
 & M & M (h.r.) & L & L (h.r.) \\
\hline
refinement level & $6$ & $7$ & $6$ & $7$ \\
$\Delta x$ [pc] & $1.02$ & $0.51$ & $0.63$ & $0.31$ \\
$\langle\dot{M}\rangle$ $\left[10^{-6}~\right.$M$_\odot~$yr$\left.^{-1}\right]$ & $11.1$ & $11.2$ & $4.9$ & $5.0$ \\
$\sigma\left(\langle\dot{M}\rangle\right)$ $\left[10^{-6}~\right.$M$_\odot~$yr$\left.^{-1}\right]$ & $1.1$ & $9.8$ & $0.3$ & $3.4$ \\
$\langle\Phi\rangle$ $\left[10^{-10}~\right.$M$_\odot~$yr$^{-1}~$pc$\left.^{-2}\right]$ & $3.2$ & $3.7$ & $4.4$ & $5.1$ \\
\hline
\end{tabular}
\end{table}
\begin{figure*}
\centering
\includegraphics[width=.75\textwidth]{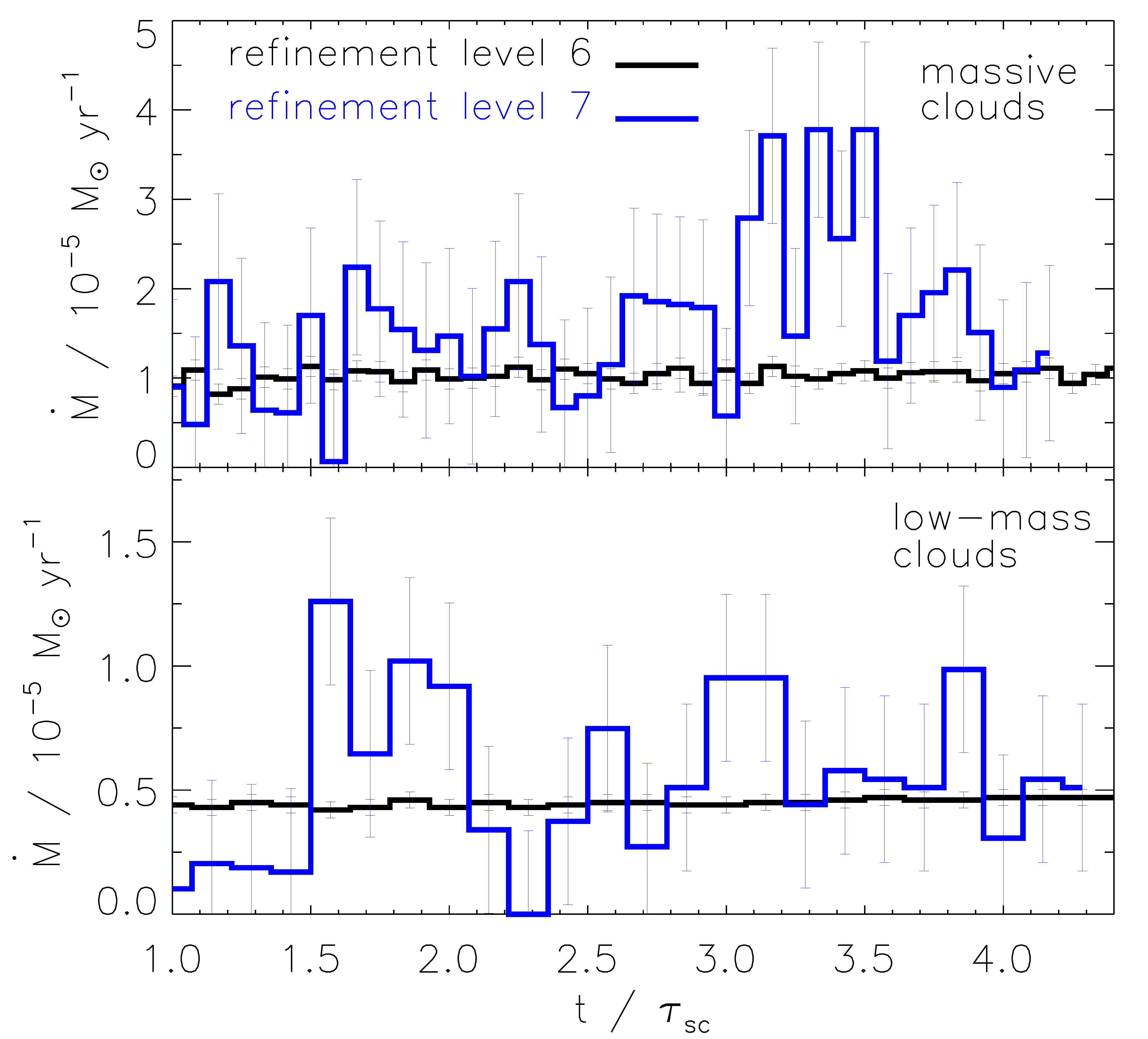}
\caption{\emph{Upper plot:} Mass-transfer rates of model M (\emph{black line}) and its highly resolved variant (\emph{blue line}). \emph{Lower plot:} The same for model L.}
\label{fig:resolution}
\end{figure*}
%
\bsp	
\label{lastpage}
\end{document}